\pdfoutput=1
\documentclass[aps,prl,twocolumn, superscriptaddress]{revtex4}
\usepackage[utf8]{inputenc}
\usepackage[T1]{fontenc}
 \usepackage{lmodern}
\usepackage{mathptmx} 
\usepackage{amsmath,amssymb}
\usepackage{xcolor}
\usepackage{graphicx}
\graphicspath{{/}{SM/}}
\usepackage[colorlinks,linkcolor=blue,urlcolor=blue,citecolor=blue]{hyperref}
\usepackage[normalem]{ulem}
\usepackage{xspace}

\usepackage{fancyhdr}
\newcommand{\eref}[1]{Eq.~(\ref{#1})}%
\newcommand{\Eref}[1]{Equation~(\ref{#1})}%
\newcommand{\fref}[1]{Fig.~\ref{#1}} %
\newcommand{\Fref}[1]{Figure~\ref{#1}}%

\begin{document}

\title{Freezing Transition in the Barrier Crossing Rate of a Diffusing Particle}

\author{Sanjib Sabhapandit}
\affiliation{Raman Research Institute, Bangalore 560080, India}
\author{Satya N. Majumdar}
\affiliation{LPTMS, CNRS, Universit\'e Paris-Sud, Universit\'e Paris-Saclay, 91405 Orsay, France}
\date[]{Published 9 November 2020. DOI: \href{https://doi.org/10.1103/PhysRevLett.125.200601}{10.1103/PhysRevLett.125.20060}}

\begin{abstract}

We study the decay rate $\theta(a)$ that characterizes the late time exponential decay of the 
first-passage probability density $F_a(t|0) \sim e^{-\theta(a)\, t}$ of a diffusing particle in a one 
dimensional confining potential $U(x)$, starting from the origin, to a position located at $a>0$.  For 
general confining potential $U(x)$ we show that $\theta(a)$, a measure of the barrier (located at $a$) 
crossing rate, has three distinct behaviors as a function 
of $a$, depending on the tail of $U(x)$ as $x\to -\infty$. 
In particular, for potentials behaving as $U(x)\sim |x|$ when $x\to -\infty$,
we show that a novel freezing transition occurs at a critical value $a=a_c$, i.e, 
$\theta(a)$ increases monotonically as $a$ decreases till $a_c$, and for $a
\le a_c$ it freezes to $\theta (a)=\theta(a_c)$. 
Our results are established using a general mapping to a quantum problem and by exact
solution in three representative cases, supported by numerical simulations.
We show that the freezing transition 
occurs when in the associated quantum problem, the gap between the 
ground state (bound) and the continuum of scattering states vanishes.

\end{abstract}

\maketitle

\thispagestyle{fancy}

Consider an overdamped Brownian particle on a line in the presence of an external potential $U(x)$, whose position 
$x(t)$ evolves by the Langevin equation
\begin{equation}
\frac{dx}{dt} = - \frac{1}{\Gamma}\, U'(x) + \sqrt{2 D} \, \eta(t), 
\label{LE}
\end{equation}
where $D=k_B T/\Gamma $, with $k_B$, $T$, and $\Gamma$ being the Boltzmann constant, temperature and 
the friction coefficient respectively.  The white noise $\eta(t)$ has zero mean and is $\delta$ correlated: $\langle 
\eta(t) \rangle =0$ and $\langle \eta(t) \eta(t')=\delta(t-t')$. 
For a particle starting at a local minimum $x_0$ of the 
potential $U(x)$, what is the rate $\kappa (a)$ with which the particle crosses over 
a barrier of relative height 
$\Delta U = U(a)-U(x_0)$, located at $a > x_0$? Estimating $\kappa (a)$ is one of the most important and celebrated 
problems in the theory of reaction kinetics,
often known as Kramers problem.
It has found immense applications in physics, chemistry, biology and engineering sciences (for a review with nice 
historical aspects see~\cite{HTB1990}).  Assuming near-equilibrium position distribution inside the potential well, the 
escape rate can be estimated by computing the flux across the barrier~\cite{Farkas1927, KS1934, BD1935, Kramers1940, 
vanKampen, HTB1990}. In the low temperature and/or large barrier limit, it is well approximated by the 
van't Hoff-Arrhenius 
form~\cite{Hoff1884, Arrhenius1889}
\begin{math}
\kappa(a) \sim e^{-\Delta U/(k_B T)}.
\end{math}

Another alternative approach~\cite{PAV1933}, that even predates Kramers, 
consists in estimating $1/\kappa(a)$ by the mean first-passage time 
$T_a(x_0)$ from $x_0$ to $a$. A quantity that carries more information 
is the full distribution $F_a(t| x_0)$ of the first-passage time to 
level $a$ starting at $x_0$. Evidently, $T_a(x_0)$ is just the first 
moment of the distribution. The cumulative first-passage distribution 
$S_a(t|x_0)=\int_t^\infty F_a(t'|x_0)\, dt'$ is known as the survival 
probability, which can in principle be computed by solving the 
Fokker-Planck equation for the probability density with an absorbing 
boundary condition at 
$x=a$~\cite{Risken1984,SM1999,Redner2001,SM2005,BMS2013}.  For a 
confining potential $U(x)$, usually the Fokker-Planck operator has a 
discrete spectra, and hence the survival probability [and consequently 
the first-passage probability] is expected to decay exponentially at 
late times: $S_a(t|x_0) \sim e^{-\theta (a) \,t}$ where the decay rate 
$\theta(a)$ gives another estimate of the escape rate $\kappa(a)$. While 
the mean first-passage time $T_a(x_0)$ can be computed explicitly for 
arbitrary potential $U(x)$~\cite{Risken1984}, the decay rate $\theta(a)$ 
is much harder to compute and there is no known formula for $\theta(a)$ 
for general potential $U(x)$, though there has been recent progress for specific cases~\cite{GM2016a, HG2018, HG2019, HG20192, MKC2019}.
 
We remark that even though the problem is posed here 
in the language
of barrier crossing, the first-passage probability  of a diffusing particle
in a confining potential has a much broader applicability ranging
from search processes in animal foraging for food~\cite{BLMV2011,VDRS2011}, 
all the way to gene transcription regulation~\cite{GM2016b}. 
For example, in the context
of foraging, an animal is typically confined in its
home range and searches for a target (food) located at a distance $a$
(from its nest)
which {\it need not be always large}, as in the Kramers problem. Hence,
estimating $\theta(a)$ for all $a$ is a fundamental problem of
broad interest.

\begin{figure}
\includegraphics[width=\hsize]{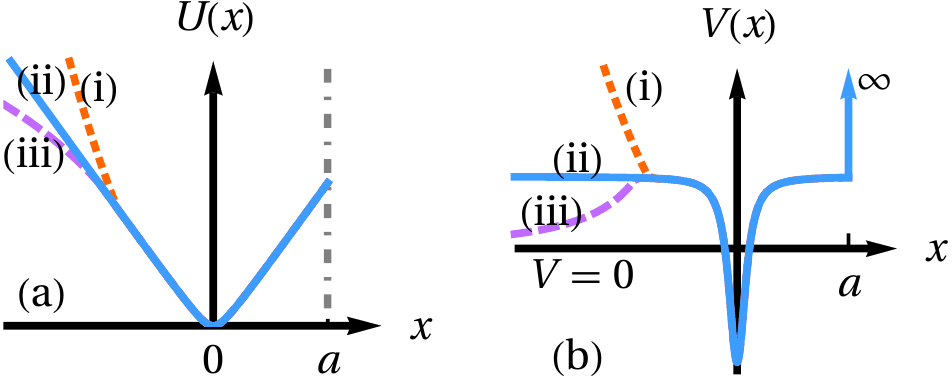}
\caption{\label{potentials}  (a) A schematic illustration of classical potentials $U(x)$ whose left tails 
as $x\to -\infty$, increase (i) faster than $|x|$ (red dotted line), (ii) as $|x|$ (blue solid line), and (iii) slower 
than $|x|$ (magenta dashed line). There is an absorbing barrier at $x=a$ (dot-dashed line). (b) Schematic illustrations 
of the corresponding quantum potentials $V(x)$ in \eref{V}, whose left tails as $x\to -\infty$, (i) diverges (red 
dotted line), (ii) approaches a constant (blue solid line), and (iii) tends to zero (magenta dashed line) respectively. 
$V(x)=\infty$ for $x \ge a$.}
\end{figure}

\begin{figure}
\includegraphics[width=\hsize]{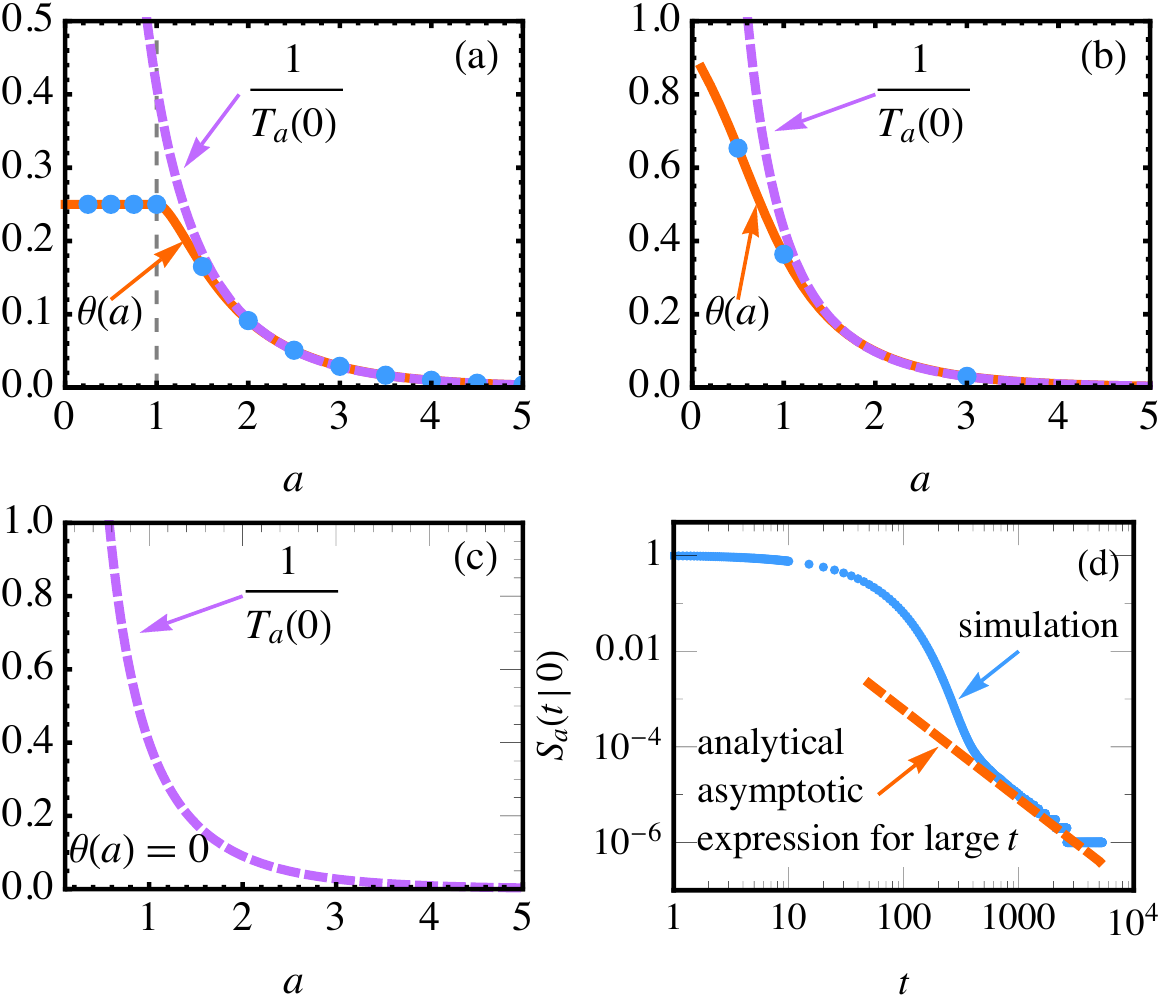}                                                                                                                                                                                                                                                                                                                                                                                                                                                                                                                                                                                                                                                                                                                                                                                                    
\caption{\label{theta-a}  (a) The (red) solid line plots the
analytical $\theta(a)$ vs. $a$ in \eref{theta-a-modx}, for the potential in \eref{potential2} 
in case (ii), with $\alpha=D=1$. The (blue) points represent the simulation
results. The (magenta) dashed line plots the analytical
expression (see \cite{SM}) of the inverse of the mean first-passage time.
The vertical (gray) dashed line marks $a=a_c$. (b) Same plot as in (a) but for
the case (i) in \eref{potential2} with $\mu=\alpha=D=1$ and $b=2$. 
(c) The (magenta) dashed line plots the  inverse of the mean first-passage time 
for case (iii) in \eref{potential2}, with $\alpha=D=\lambda=1$ and $b=c=e$. In this case 
$\theta(a)=0$ for all $a$. (d) The (blue) points represent simulation results  
for the survival probability $S_a(t|0)$ for the 
same potential as in (c), while the (red) dashed line represents 
the asymptotic decay of $S_a(t|0)$ 
in \eref{S-long-t-log}.
}
\end{figure}

For large $a$, the three estimates of $\kappa(a)$, namely, the Kramers estimate, the inverse mean first-passage time 
$1/T_a(x_0)$, and the decay rate $\theta(a)$, all have the Arrhenius 
form $\sim e^{-U(a)/(k_B T)}$  [see \cite{HG20192} for a more refined
estimate of $\theta(a)$ for large $a$]. However, 
in many  search processes discussed above, the location $a$ of 
the barrier (or the target) is not necessarily large and
the three measures may have different $a$ dependences,
in particular for small $a$. Indeed, the discrepancy between
the last two measures for small $a$ is expected for compact
diffusion~\cite{GM2016b}, which is the case here since the potential is confining.
In this Letter, we study the
$a$ dependence of the three measures for general potential $U(x)$ and show
that indeed for small $a$, they are quite different from each other. In particular,
we show that $\theta(a)$ for general $U(x)$ displays a rich and robust $a$ dependence,
depending on the tail of $U(x)$ as $x\to -\infty$ (see \fref{potentials}), that is not captured
by the other two measures of $\kappa(a)$. More precisely, we find 
(i) if $U(x)$ increases faster than $|x|$ as $x\to -\infty$, then $\theta(a)$ increases 
monotonically as $a$ decreases, (ii) if $U(x) \sim |x|$ as $x\to -\infty$, then there is a critical value of $a$ at 
$a=a_c$, where a novel {\it freezing transition} occurs, i.e., $\theta(a)$ increases 
monotonically as $a$ decreases till 
$a_c$, and for $a \le a_c$ the decay rate $\theta (a)=\theta(a_c)$, and (iii) if $U(x)$ increases slower than $|x|$ as 
$x\to -\infty$, then $\theta(a)=0$ for all $a$, indicating a slower than exponential decay with time, of the 
first-passage probability (see \fref{theta-a}).

\begin{figure}
\includegraphics[width=.9\hsize]{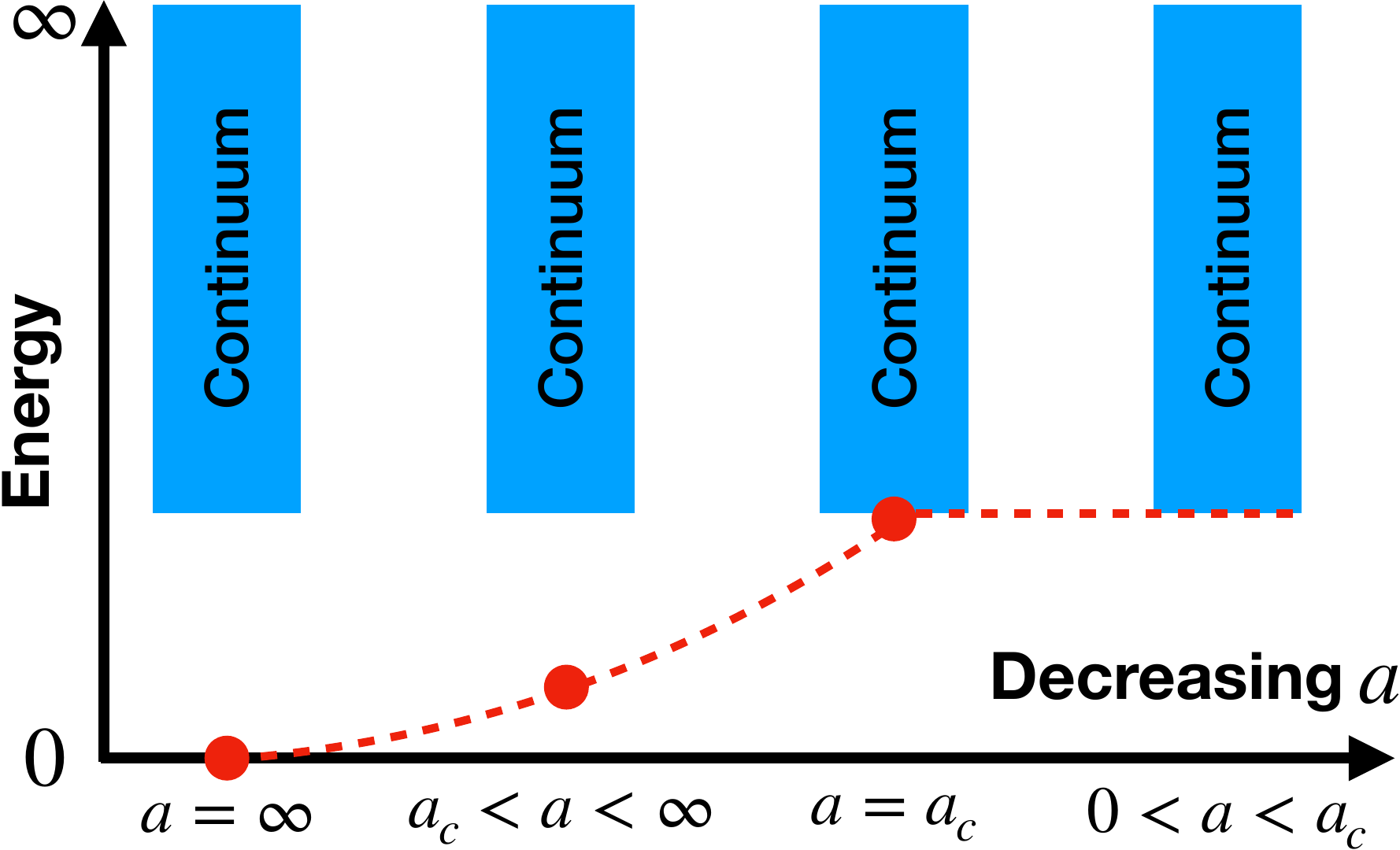}
\caption{\label{spectrum-fig}  Schematic diagram of energy levels for the quantum potential 
$V(x)=\alpha^2/(4 D) -\alpha \delta(x)$ for $x <a$ and $V(x)=\infty$ for $x \ge a$, for different values of $a$. The 
(red) dots represent the ground state energy for different values of $a$ whereas the (blue) bands represent the 
continuum of energy levels from $\alpha^2/(4D)$ to $\infty$. The gap vanishes at $a=a_c=D/\alpha$.}
\end{figure}

We establish this behavior by mapping to a quantum problem, 
where the quantum potential (see \fref{potentials}) has always bound states in case 
(i), while in case (iii) it only has 
a continuous spectrum of scattering states. In the borderline case (ii), the spectrum has a bound 
state separated by a 
gap from the continuum of scattering states for $a > a_c$, and the gap vanishes as $a\to a_c^+$ (see
\fref{spectrum-fig}). 
In case (i) and case 
(ii) with $a > a_c$, where the spectrum has bound states, 
$\theta(a)$ coincides exactly with
the ground state energy of the quantum problem. 
The inverse mean first-passage time $1/T_a(x_0)$, 
in contrast, always increases 
monotonically with decreasing $a$, and hence misses this novel freezing transition 
at $a=a_c$ (see \fref{theta-a}).  This transition is
also consistent with the powerful interlacing theorem derived
in \cite{HG2018,HG2019}; for details see the Supplemental Material~ \cite{SM}.
The mapping to the quantum problem makes it evident that the scenario presented above 
holds generically for any confining potential 
$U(x)$. In addition, we show 
the validity of this generic behavior by explicit exact solution in three 
representative cases here (for another example, see ~\cite{SM}).

We start with the Fokker-Planck equation for the probability density function  $P(x,t)$ of the particle to be at 
$x$ at time $t$, without having crossed the level at $x=a$,
\begin{equation}
\frac{\partial P}{\partial t}= D \frac{\partial^2 P}{\partial x^2} +  
\frac{\partial}{\partial x} \left[U'(x) P \right], 
\label{FP.1}
\end{equation}
where we set $\Gamma=1$ for simplicity. \Eref{FP.1} holds in $x\in (-\infty, a)$ with an absorbing 
boundary condition $P(a,t)=0$ at $x=a$ and also, $P(x\to-\infty,t)=0$. 
 We assume that the particle 
starts at the origin at $t=0$, i.e., $P(x,0)=\delta(x)$ and the barrier location
$a\ge 0$ to be on the right of the initial position of the particle. If $a<0$, one can
reverse $x$ and perform a similar analysis.
With the 
transformation~\cite{Risken1984}, 
\begin{math}
P(x,t) = e^{-[U(x)-U(0)]/(2D)} \, \psi (x,t),
\end{math}
\eref{FP.1} gets mapped to the time-dependent Schr\"odinger equation in imaginary time, 
\begin{math}
-\partial \psi/\partial t=\mathcal{H} \psi(x,t),
\end{math}
where
\begin{math}
\mathcal{H}=-D (\partial^2/\partial x^2) + V(x), 
\end{math}
with the initial condition $\psi(x,0)=\delta(x)$ and the quantum potential
\begin{equation}
V(x)=\frac{[U'(x)]^2}{4D} -\frac{U''(x)}{2},\quad\text{for} ~x<a.
\label{V}
\end{equation}
Here we assume $U(x)$ to be twice differentiable 
almost everywhere.  
The absorbing boundary condition $P(x=a,t)=0$ translates into $\psi(x=a,t)=0$, which in the quantum 
problem,  corresponds to having an infinite barrier at $x=a$, i.e., $V(x)=\infty$ for $x \ge a$. The wave function $\psi(x,t)$ can be 
written in the eigenbasis of $\mathcal{H}$, as
\begin{equation}
\psi(x,t) = \sum_E \phi_E^*(0)\, \phi_E(x)\, e^{-E t}
\label{spectral}
\end{equation} 
where $\mathcal{H} \phi_E(x) = E \phi_E(x)$, and we have used $\psi(x,0)=\delta(x)$. The sum over $E$ includes both the
discrete and the continuous part of the spectrum. When the ground state is bound and is separated by a finite gap from the 
rest of the spectrum, then it follows  from  \eref{spectral}, that at late times $P(x,t) \sim 
e^{-E_0 t}$ where $E_0$ is the ground state energy. Correspondingly the survival probability $S_a(t|0) = 
\int_{-\infty}^a P(x,t)\, dx \sim e^{-\theta(a)\, t}$ with $\theta(a) = E_0$.

Since there is an infinite barrier at $x=a$, whether the Hamiltonian $\mathcal{H}$ has a bound state or not depends only 
on the behavior of $V(x)$ as $x\to -\infty$. For example, if the classical potential $U(x)$ increases faster than $|x|$ 
as $x\to -\infty$ such as $U(x)\sim (-x)^\gamma$ with $\gamma >1$, then it is easy to see from \eref{V} that $V(x) \sim 
(-x)^{2(\gamma-1)}$, and hence, $V(x)$ diverges as $x\to-\infty$. In this case, clearly, the quantum problem will have 
only bound states. On the other hand, if $\gamma <1$, then $V(x)\to 0$ as $x\to -\infty$, indicating that the quantum 
problem will only have scattering states. In the marginal case, $\gamma=1$, $V(x)$ approaches a constant as $x\to 
-\infty$, and in this case, one would expect that a bound state may or may not exist depending on the value of $a$. To 
illustrate this general scenario, we present below an exact solution for a representative $U(x)$ whose $x\to-\infty$ tails can be 
tuned as in the three cases above. More precisely, we choose
\begin{subequations}
\label{potential}
\begin{equation}
U(x)=\alpha |x| \quad \text{for} ~ -b < x< a,
\label{potential1}
\end{equation}
with $b >0$, and for $x <-b$,
\begin{equation}
U(x) = \begin{cases}
\frac{1}{2}\, \mu x^2 & \text{for case (i)}\\
\alpha |x|  & \text{for case (ii)}\\
c\ln (-x/\lambda) & \text{for case (iii)}\\
\end{cases}
\label{potential2}
\end{equation}
\end{subequations}
where $\lambda >0$ is a length scale and for the sake of continuity of the potential at $x=-b$, we set $\mu=2\alpha/b$ and 
$c\ln (b/\lambda)=\alpha b$ (see \fref{potentials}).

For convenience, we start our analysis for the marginal case (ii) in \eref{potential2}
where $U(x)= \alpha\, |x|$ for $x<a$ and
show explicitly that a freezing transition occurs at 
the critical value $a=a_c=D/\alpha$. 
The cases (i) and (iii) will be discussed subsequently. 
The quantum potential $V(x)$ from \eref{V} is 
then given by $V(x)=\alpha^2/(4 D) -\alpha \delta(x)$ for $x< a$ 
and $V(x)=\infty$ for $x\ge a$. In this case, the Schr\"odinger equation  can be solved either by 
spectral decomposition as in \eref{spectral} or equivalently by taking the  Laplace transform  of the equation with respect 
to $t$. In the later case, the spectral values of $E$ manifest as poles (for the discrete part of the spectrum) or as a 
branch cut (for the continuous part of the spectrum).   Skipping details (see~\cite{SM}),  the solution in the Laplace space $\tilde{\psi}(x,s)=\int_0^\infty 
\psi(x,t)\, e^{-st}\, dt$,  reads
\begin{equation}
\tilde{\psi}(x,s) = \begin{cases}
\frac{1}{A(p)}\, \bigl[1-e^{-pa/D} \bigr]\, e^{p x/(2D)} &\text{for}~ x\le 0,\\[2mm]
\frac{1}{A(p)}\,\bigl[1 - e^{-p (a-x)/D} \bigr]\, e^{-p x/(2D)} &\text{for}~ 0 \le x \le a,\\
\end{cases}
\label{psi-laplace2}
\end{equation}
where
\begin{equation}
A(p)=p-\alpha \left(1-e^{-pa/D}\right)
\quad\text{with}~~
 p=\sqrt{\alpha^2+4 D s}. 
\label{Ap}
\end{equation}
It is evident from Eqs.~\eqref{psi-laplace2} and \eqref{Ap} that there is a branch cut at 
$s=-\alpha^2/(4D)$, signaling 
a continuum of eigenstates with energy $E \ge \alpha^2/(4D)$ [see \fref{spectrum-fig}]. In addition, for $a> 
a_c=D/\alpha$, there is an isolated pole at $s=s^*(a) =-(\alpha^2 -p^{*2})/(4D)$ where $0 < p^*(a) < \alpha $ is the 
nonzero solution of the transcendental equation $A(p^*)=0$ (see \cite{SM}), where $A(p)$ is given in 
\eref{Ap}. This corresponds to a bound state (which is indeed the ground state) with energy $E_0(a)=-s^*(a)$. Thus there 
is a gap in the spectrum $\Delta(a)=\alpha^2/(4D) -E_0(a) =p^{*2}/(4D)$, between the ground state and 
the excited 
states. Consequently, the survival and the first-passage probabilities decay as $\sim e^{-\theta(a) t}$ for large $t$ where 
$\theta(a)=E_0(a)$. As $a\to a_c^+$, the gap vanishes as $\Delta(a)\sim (a-a_c)^2$ (see~\cite{SM}). For $a \le a_c$, the 
spectrum has only a continuous part consisting of scattering states with $E \ge \alpha^2/(4D)$. By analyzing the Laplace 
transform~\cite{SM} we find that the survival probability decays as 
$S_a(t|0) \sim t^{-3/2}\, e^{-\alpha^2\, t/(4D)}$ for $a 
< a_c$ and $S_a(t|0) \sim t^{-1/2}\, e^{-\alpha^2\, t/(4D)}$ exactly at $a=a_c$.  
Hence, $\theta(a) 
= -\lim_{t\to\infty}\, t^{-1}\, S_a(t|0)$ is given by
\begin{equation}
\theta(a) =\begin{cases}
(\alpha^2-p^{*2})/(4D) & \text{for}~ a> a_c =D/\alpha\\
\alpha^2/(4D) &\text{for}~ a \le a_c.
\end{cases}
\label{theta-a-modx}
\end{equation}
The freezing value $\theta(a_c)=\alpha^2/{4D}$ also predicts,
using the interlacing theorem~\cite{HG2018}, 
where the continuum part of the relaxation spectrum starts~\cite{SM}.
In contrast, the inverse mean first-passage time $1/T_a(0)$ increases monotonically with decreasing 
$a$ (see \cite{SM}).  
Interestingly, a similar nonmonotonic behavior of $\theta(a)$ was recently observed in the 
context of the dry friction problem~\cite{MKC2019}. In \fref{theta-a}~(a), we plot our analytical
expression \eref{theta-a-modx} and compare it with numerical simulations performed for few values of $a$, finding
excellent agreement. For comparison, we also plot $1/T_a(0)$ (shown by the dashed line in
\fref{theta-a}), which increases monotonically with decreasing $a$. While this result
is proved here for the specific potential \eref{potential2}, it is clear from the
general mapping to the quantum problem that this freezing transition is robust as long
as $U(x) \sim |x|$ as $x\to -\infty$ and its existence should
not depend on the details of $U(x)$ in the bulk. We demonstrated this for
another choice of the potential $U(x)$ in the Supplemental Material~\cite{SM}.

We now turn to the cases (i) and (iii) in \eref{potential2}.
While calculations in these two cases can also be carried out by mapping to the 
quantum problem (which indeed helps understanding the physics better), computationally 
it turns out to be more convenient to use
a shorter backward Fokker-Planck approach~\cite{Risken1984,SM2005,BMS2013} for
the survival probability $S_a(t|x_0)$ where
the starting position $x_0$ is treated as a variable. 
The first-passage probability is then derived from
the relation $F_a(t|x_0)=-\partial_t S_a(t|x_0)$. 
The backward Fokker-Planck equation reads
\begin{equation}
\frac{\partial S_a}{\partial t} = D \frac{\partial^2 S_a}{\partial x_0^2} -  U'(x_0)\, \frac{\partial S_a}{\partial x_0}
\label{S_a}
\end{equation} 
with the initial condition $S_a(0|x_0)=1$ and the boundary conditions, 
$S_a(t|x_0\to -\infty)=1$ and $S_a(t|x_0=a)=0$. 
Skipping details (see \cite{SM}), the solution in the Laplace space  $\tilde{S}_a(s|x_0)= \int_0^\infty S_a(t|x_0)\, e^{-st}\, dt$,  reads
\begin{equation}
\tilde{S}_a (s|0) = \frac{1}{s} \bigl[1-\tilde{F}_a(s|0)\bigr]
\end{equation}
where the Laplace transform of the first-passage time distribution is given by
\begin{equation}
\tilde{F}_a(s|0)=\frac{p}{B(s)}\, e^{-(\alpha+p) a/(2D)}\, \chi(s),
\label{tildeF_a}
\end{equation}
with $p=\sqrt{\alpha^2+4 D s}$ and the expressions of $B(s)$ and $\chi(s)$---which are different in 
the cases (i) 
and (iii)---are given in \cite{SM}. Analyzing $\tilde{F}_a(s|0)$, we find that 
$s=-\alpha^2/(4D)$ is no longer a branch point.

In case (i), where $U(x) \sim x^2$ as $x\to -\infty$, the quantum potential $V(x)$ 
in \eref{V} also diverges  
$\sim x^2$ as $x\to -\infty$.  Hence, the quantum problem has only bound states 
with discrete spectrum. By analyzing (see \cite{SM}) \eref{tildeF_a}, we indeed find that the denominator 
$B(s)$ has infinite number of 
zeros---equivalently, $\tilde{F}_a(s|0)$ has infinite number of poles---on the negative line $-\infty <s<0$. This
infinite set of poles $ -\infty < \dotsb< s_2^*(a) < s_1^*(a) < s_0^*(a) < 0$ corresponds to having only bound states 
with discrete energy levels $E_i(a)=-s_i^*(a)$ with $i=0,1,\dotsc,\infty$. Therefore, both the first-passage and the 
survival probability decays as $F_a(t|0) \sim S_a(t|0) \sim e^{-\theta(a) t}$, where $\theta(a) = E_0(a) = -s_0^*(a)$. 
\Fref{theta-a}~(b) plots $\theta(a)$ as a function of $a$ together with $1/T_a(0)$.

Turning now to case (iii), where $U(x) \sim c\,\ln (-x)$ as $x\to -\infty$ in \eref{potential2},
we chose, for simplicity, $\lambda=1$ and $c>D$. The latter condition
ensures that in the absence of the absorbing wall at $a$, the
stationary Boltzmann distribution
$P(x, t\to \infty) \propto e^{-U(x)/D}$ is normalizable, i.e., $\int e^{-U(x)/D}\, dx$ is finite. 
Diffusion in such potentials with logarithmic tails has been studied extensively in various contexts
such as in the denaturation process of DNA molecules~\cite{BKM2009}, momentum distribution of cold atoms in
optical lattices~\cite{CDC1991,MEZ1996,KB2010}, among many 
others~\cite{EH2005,LMS2005,CDR2009,Bray2000,Lo2002,HMS2011,RR2020}. 
In this case, the associated quantum potential $V(x)\to 0$ as 
$x\to -\infty$ from \eref{V}.  Therefore, the quantum problem has only scattering states and no bound state. Indeed we find 
that $\tilde{F}_a(s|0)$ does not have any pole, even when $a\to\infty$.  Anticipating the scattering states to 
lead to a power-law decay for $F_a(t|0)$ at late times, we analyze
$\tilde{F}_a(s|0)$ near $s=0$, and from it deduce the asymptotic decay
of $F_a(t|0)$ for large $t$. 
For a noninteger $\nu = (1+c/D)/2 \in (n,n+1)$ with $n\ge 1$ being an integer, 
we find the following late time decay~\cite{SM}
\begin{equation}
F_a(t|0)  = \nu\,  a_\nu\, t^{-(\nu+1)} + o( t^{-(\nu+1)}).
\label{fpdecay.1}
\end{equation}
where the amplitude $a_\nu$ can be computed explicitly~\cite{SM}.
Consequently, the survival probability decays as
\begin{equation}
S_a(t|0) = \int_t^\infty F_a(t'|0)\, dt' =  a_\nu\, t^{-\nu} + o(t^{-\nu}). 
\label{S-long-t-log}
\end{equation}
For integer values of $\nu$, there are additional $\ln t$ corrections.
Hence, $\theta(a) = -\lim_{t\to\infty}\, t^{-1}\, 
\ln S_a(t|0)$ is zero for all $a$, while $1/T_a(0)$ is still nonzero and a 
monotonic function of $a$ [see \fref{theta-a}~(c)]. In \fref{theta-a}~(d), we verify the analytical 
prediction in \eref{S-long-t-log} by numerical simulation.

In conclusion, our main result is that $\theta(a)$, as a 
function of decreasing target location $a$, has substantially different 
behaviors depending on the far negative tail of the confining potential 
$U(x)$. {\em Far negative tail} actually means $|x|\gg \xi$ for negative 
$x$, where $\xi$ denotes the typical width of the confining potential 
near its minimum, e.g., in the context of animal foraging for food, 
$\xi$ denotes the size of the home range. At first sight, this may look 
puzzling: why does the far negative tail of $U(x)$ affect $\theta(a)$ 
for $a\sim O(1)>0$? Qualitatively, if the potential is sufficiently 
confining [$U(x)\sim |x|^{\gamma}$ as $-x \gg \xi$ with $\gamma>1$], the 
typical trajectory remains confined and the particle feels the presence 
of the absorbing barrier located at $a$ more strongly. However, for $\gamma\le 1$ 
(where the spectrum of the associated quantum problem has scattering 
states), the typical trajectory wanders off to the far negative side 
(since the potential is not sufficiently confining on that side), and 
hence the particle becomes more insensitive to the presence of the 
absorbing barrier at $a>0$. In this Letter, this qualitative 
understanding is made precise via the exact analysis of the associated 
quantum problem. This led to a surprising  ``freezing'' transition of 
$\theta(a)$ at a critical value $a_c$ for $\gamma=1$. We expect this 
freezing transition to be robust, e.g., it will hold for finite systems 
when the system size $L\gg \xi$. Finally, the potentials discussed in 
the Letter can be tailored by using holographic optical tweezers  and 
confining the movements of colloidal particles to a 
quasi-one-dimensional line using a microfluidic device~\cite{CGCKT20}, 
leading to a possible experimental measurement of $\theta(a)$.

S. N. M. acknowledges
the support from the Science and Engineering Research Board
(SERB, government of India), under the VAJRA faculty
scheme  (No.  VJR/2017/000110)  during  a  visit  to  Raman Research Institute, where part of this work was carried  
out.

\end{document}


\title{Freezing Transition in the Barrier Crossing Rate of a Diffusing Particle: Supplemental Material}

\author{Sanjib Sabhapandit}
\affiliation{Raman Research Institute, Bangalore 560080, India}
\author{Satya N. Majumdar}
\affiliation{LPTMS, CNRS, Univ. Paris-Sud, Universit\'e Paris-Saclay, 91405 Orsay, France}
\date{\today}

\begin{abstract}
We give principal details of the calculations and simulations described in the main text of the Letter.
\end{abstract}

\maketitle

\thispagestyle{fancy}

\section{The details of computing $\tilde{\psi}(x,s)$  and the survival probability for the potential $U(x)=\alpha |x|$} 
\label{S-I}

In the presence of an absorbing barrier at $x=a$, the quantum potential is given by
\begin{equation}
V(x)=\begin{cases}\displaystyle
\frac{\alpha^2}{4 D} -\alpha \delta(x) &\text{for}~ x< a,\\
\infty &\text{for}~ x\ge a.
\end{cases}
\end{equation}
Therefore, the Schr\"odinger equation (in imaginary time) in the 
in the Laplace space, $\tilde{\psi}(x,s)=\int_0^\infty 
\psi(x,t)\, e^{-st}\, dt$,  reads 
\begin{equation}
D\tilde{\psi}''(x,s) -[\alpha^2/(4 D) -\alpha \delta(x)+s] \tilde{\psi}(x,s)=-\delta(x),
\quad \text{for} ~~ x \le a,
\label{SE-Laplace-SM}
\end{equation}
with boundary conditions $\tilde{\psi}(x\to-\infty,s) =0$ and $\tilde{\psi}(x=a,s)=0$.

We solve \eref{SE-Laplace-SM} 
separately for $x < 0$ and $0 < x < a$, and then match the solutions at $x=0$, where the wave function is continuous, but 
its first derivatives undergoes a jump due to the $\delta$-function at $x=0$. 
In each of these regions
\eref{SE-Laplace-SM} reads $D\tilde{\psi}''(x,s) -[\alpha^2/(4 D) +s] \tilde{\psi}(x,s)=0$. 
Therefore, the solutions can be written as
\begin{equation}
\tilde{\psi}(x,s) = \begin{cases}
A_1 e^{px/(2D)} + B_1 e^{-px/(2D)}  &\text{for}~ x <0,\\
A_2 e^{px/(2D)} + B_2 e^{-px/(2D)}  &\text{for}~ 0 < x <a,
\end{cases}
\end{equation}
with $p=\sqrt{\alpha^2+4 Ds}$ and $A_1$, $B_1$, $A_2$ and $B_2$ are four unknown constants to be fixed from
the boundary and the matching conditions.

The boundary condition $\tilde{\psi}(x\to-\infty,s) =0$ implies $B_1=0$. 
On the other hand the absorbing boundary condition $\tilde{\psi}(x=a,s)=0$ yields 
 \begin{math}
 A_2= - B_2 e^{-p a/D} \, .
 \end{math}
Therefore, the solutions read
\begin{equation}
\tilde{\psi}(x,s) = \begin{cases}
A_1 e^{px/(2D)}  &\text{for}~ x <0,\\
B_2 \left[ 1- e^{- p(a-x)/D}  \right] e^{-px/(2D)}  &\text{for}~ 0 < x <a.
\end{cases}
\end{equation}

The two remaining constants $A_1$ and $B_2$ can be determined by matching the solutions at $x=0$. Integrating \eref{SE-Laplace-SM} in an infinitesimal region across $x=0$ gives the continuity of the solution $\tilde{\psi} (0^+,s) =  \tilde{\psi}(0^-,s) \equiv \tilde{\psi}(0,s)$ and the discontinuity of the derivatives $D \bigl[\tilde{\psi}'(0^+,s)  - \tilde{\psi}'(0^-,s)  \bigr] +\alpha \tilde{\psi}(0,s) = -1$. 
The continuity of the solution implies $A_1= B_2\bigl(1-e^{-p a/D}\bigr)$. 
Finally, the discontinuity of the solution gives $B_2 = 1/A(p)$ where 
\begin{equation}
A(p) = p -\alpha \bigl( 1-e^{-p a/D}\bigr)\, , 
\label{Ap.1}
\end{equation}
with $p=\sqrt{\alpha^2+4 Ds}$.
Therefore,
\begin{equation}
\tilde{\psi}(x,s) = \begin{cases}
\frac{1}{A(p)}\, \bigl[1-e^{-pa/D} \bigr]\, e^{p x/(2D)} &\text{for}~ x\le 0,\\[2mm]
\frac{1}{A(p)}\,\bigl[1 - e^{-p (a-x)/D} \bigr]\, e^{-p x/(2D)} &\text{for}~ 0 \le x \le a,\\
\end{cases}
\label{psi-laplace2-SM}
\end{equation}
as mentioned in the main text.

We can obtain $\psi(x,t)$ by using the Bromwich integral
\begin{equation}
\psi(x,t)=\frac{1}{2\pi i}\int_{c-i\infty}^{c+i\infty} \tilde{\psi}(x,s) \, e^{st}\, ds,
\label{eq48.1}
\end{equation}
where $c$ is a real number such that all the singularities of $\tilde{P}(x,s)$ are on the left of the 
vertical contour $\mathrm{Re}(s)=c$, in the complex-$s$ plane. The most dominant large $t$ behavior comes 
from the singularity closest to the contour, the second dominant contribution comes from the next 
singularity and so on.

\begin{figure}
\includegraphics[width=.5\hsize]{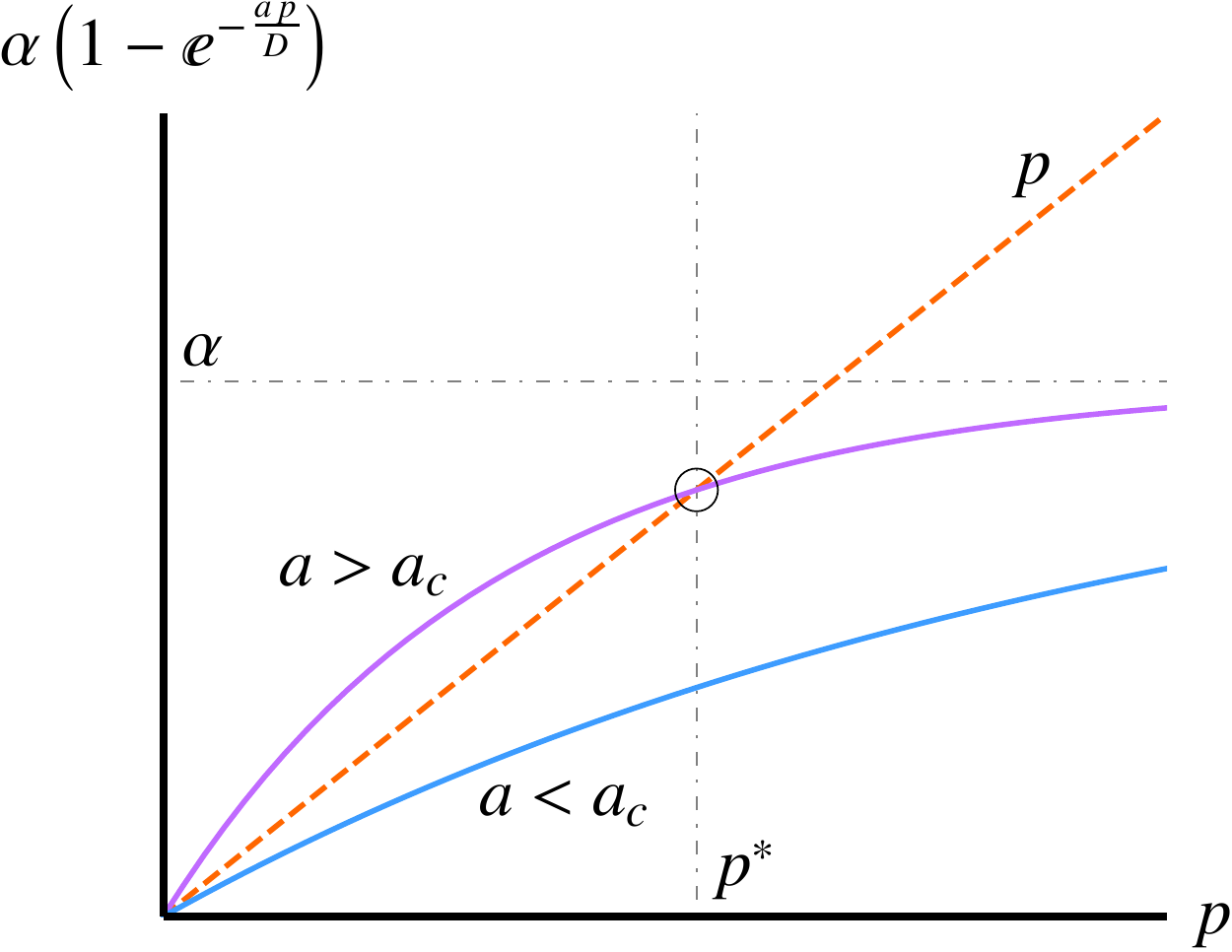}
\caption{\label{p-eqn2-fig} Graphical representation of the non-trivial (non-zero) solution $p^*$ of the transcedental equation $p=\alpha (1-e^{-p a/D})$, which exists only for $a >a_c=D/\alpha$. For $a\to \infty$, $p^*=\alpha$, whereas for $a\to a_c^+$, we have $p^*\to 0$.}
\end{figure}

Let us first consider the limit $a\to\infty$, i.e., when there is no absorbing barrier. In this case,  
we have from \eqref{Ap.1},  $A(p)=p-\alpha$, and hence,  
$\tilde{\psi}(x,s)$ has a pole at $s=0$, i.e.,  $p=\alpha$.
There is also a branch-point (and not a pole) at $s=-\alpha^2/(4 D)$, i.e.,  $p=0$.  Therefore, the leading order behavior, that corresponds to the stationary state at $t\to\infty$, comes from the contribution from the pole at $s=0$ and the approach to the stationary state comes from the branch-point at $s=-\alpha^2/(4D)$. Evaluating the residue at $s=0$ gives 
\begin{equation}
\psi(x,t\to\infty) = \frac{\alpha}{2D} 
\exp\left( -\frac{\alpha |x|}{2D}  \right).
\label{eq48.2}
\end{equation}
Consequently, using $P(x,t) = e^{-\alpha |x|/(2 D)} \psi(x,t)$, we get the stationary distribution 
\begin{equation}
P(x,t\to\infty)\equiv p_\mathrm{ss} (x) =\frac{\alpha}{2D} \exp\left( -\frac{\alpha |x|}{D} \right).
\label{eq48.3}
\end{equation}
The contribution from the branch-point gives the approach to the stationary state as
\begin{equation}
\psi(x,t) -\psi(x,t\to\infty) = Q(x,t) \exp\left(-\frac{\alpha^2}{4D} \,t\right) 
\label{eq48.4}
\end{equation}
where $Q(x,t)$ is obtained from the integral around the branch-cut from $-\infty$ to $-\alpha^2/(4D)$, and is given by
\begin{equation}
Q(x,t)=\frac{1}{\pi} \int_0^\infty dr\,\frac{e^{-rt} }{\alpha^2+4Dr}
\biggl[
\sqrt{4Dr} \cos\left(\frac{|x|\sqrt{r}}{\sqrt{D}}\right) 
-\alpha \sin\left(\frac{|x|\sqrt{r}}{\sqrt{D}}\right)
\biggr].
\label{eq48.6}
\end{equation}
The large $t$  behaviors, for arbitrary $x$,  can be computed by expanding  $[\alpha^2 + 4Dr]^{-1}$ about $r=0$ and carrying out the above integral term by term. The leading order behavior is given by
\begin{equation}
Q(x,t)= \frac{\sqrt{D}\, e^{-x^2/(4Dt)}
}{\sqrt{\pi}\,\alpha^2\,t^{3/2}}
\left(1- \frac{x^2}{2Dt} -\frac{\alpha |x|}{2D} \right)
+O(t^{-5/2}).
\end{equation}

In terms of the energy spectrum of the quantum Hamiltonian $\mathcal{H}$ given in the Schr\"odinger 
equation in the main text, there is a single bound state corresponding to energy $E_0=0$, and continuum 
spectrum of scattering states with energy $E \ge \alpha^2/(4D)$. Hence, the gap is given by  
$\Delta=\alpha^2/(4\,D)-E_0= \alpha^2/(4D)$.  The continuum energy spectrum 
from $\alpha^2/(4\,D)$ to $\infty$, manifests as the 
branch-cut from $-\alpha^2/(4D)$ to $-\infty$ in the Laplace transform $\tilde{\psi}(x,s)$.

We now turn to finite $a$. As we bring the absorbing wall from $\infty$ to a finite value $a$, 
the location of the rightmost pole in \eref{psi-laplace2-SM} 
shifts from $s^*(\infty)=0$ to a negative value at $s=s^*(a)$ (with $s^*(a)<0$) that varies
continuously with $a$. In the quantum problem, this corresponds to the fact
that the ground state, while still remains a bound state, its energy $E_0(a)=-s^*(a)$ increases
continuously with decreasing $a$. 
On the other hand, the position of the branch-point of \eref{psi-laplace2-SM} remains fixed at 
$s=-\alpha^2/(4\,D)$, i.e., at $p=0$, irrespective of $a$.  
Therefore, the continuum spectrum for the scattering states is always from 
$\alpha^2/(4\,D)$ to $\infty$.
Clearly, as $a$ decreases, at some critical value $a=a_c$, the gap, $\Delta(a)= \alpha^2/(4\,D)-E_0(a)$, between
the scattering band starting at $\alpha^2/(4\, D)$ and the ground state $E_0(a)$
must vanish, triggering a phase transition.  

To locate this critical value $a_c$, we look for the rightmost pole
of \eref{psi-laplace2-SM}, i.e., the nonzero solution of the transcendental
equation $A(p)=p -\alpha \bigl( 1-e^{-p a/D}\bigr)=0$. In Fig. (\ref{p-eqn2-fig}) we
plot both $\alpha \bigl( 1-e^{-p a/D}\bigr)$  (solid lines) and $p$ (dashed line) vs $p$. The slope of  $\alpha \bigl( 1-e^{-p a/D}\bigr)$  at $p=0$ is given by $\alpha\, a/D$.
The two curves $p$ and  $\alpha \bigl( 1-e^{-p a/D}\bigr)$  will cross each other at a nonzero value $p*>0$ only
for the slope $\alpha\, a/D>1$, i.e., $a>a_c= D/\alpha$. As $a\to a_c$ from above, the
nonzero solution $p^*\to 0$, triggering the phase transition. In the quantum language,
at $a=a_c$, the gap $\Delta(a)$ vanishes.
The mechanism of this phase transition is thus very similar to the mean-field transition
in ferromagnetic Ising model. For $a<a_c$, $A(p)=0$ has only a trivial solution $p=0$ (which
however is not a pole as it cancels with the numerator of $\tilde{\psi}(x,s)$).
For $a > a_c=D/\alpha$, the pole at $s=s^* <0$ 
on the real-$s$ line is given by
\begin{equation}
s^*(a) =-\frac{1}{4D} \left[\alpha^2 -p^{*2} (a) \right] \, .
\label{eq60}
\end{equation}
Therefore, for $a >a_c$, the energy of the bound state is given by 
\begin{equation}
E_0(a)=-s^*(a) =\frac{1}{4D} \left[\alpha^2 -p^{*2} \right]\, ,
\label{eq61.1}
\end{equation}   
and the corresponding gap in the spectrum is given by
\begin{equation}
\Delta(a) \equiv \frac{\alpha^2}{4D} - E_0(a) = \frac{[p^{*}(a)]^2}{4D}.
\label{eq61.2}
\end{equation}
The gap vanishes at $a\to a_c^+$, as $p^*\to 0$. There is no pole, and only the branch-point at 
$s= -E_1=-\alpha^2/(4\, D)$, 
for $a<a_c$ [see \fref{p-eqn2-fig}]. Therefore, the spectrum remains gapless for $a<a_c$.

For $a>a_c$, $\psi(x,t)$ is given by
\begin{align}
\psi(x,t) = R(x) \, e^{-E_0(a)  t}  + Q(x,t) \, e^{-\alpha^2 t/(4D)} ,
\label{eq61.3}
\end{align} 
where $R(x=\lim_{s\to s^*} (s-s^*) \tilde{\psi}(x,s)$ arises from the residue at $s^*$ 
The prefactor $Q(x,t)$ of the sub-dominant  branch-point contribution is from the contour  integral around the branch-cut from $-\alpha^2/(4D)$ to $-\infty$, and the leading order time dependence is given by $Q(x,t) \sim t^{-3/2}$. 
 
At $a=a_c$,  there is no pole corresponding to $p=0$, and only a branch-point.   In this case
\begin{align}
\psi(x,t) =  Q(x,t) \, e^{-\alpha^2 t/(4D)} ,
\label{eq61.6}
\end{align} 
where the leading order time dependence of $Q(x,t)$ is given by $Q(x,t) \sim 1/\sqrt{t}$.

For $a<a_c$, again there is no pole and the contribution comes only from the branch-point. Therefore, 
$\psi(x,t)$, is still given by \eref{eq61.6}, however, with a different $Q(x,t)$. In particular, $Q(x,t) 
\sim t^{-3/2}$ for large $t$.

\vspace{0.3cm} 

{\noindent {\bf {Survival/first-passage probability:}}} The survival probability $S_a(t|0)$ with the starting position $x_0=0$,  can be obtained by integrating  $P(x,t)$ obtained above,  over $x$ fro $0-\infty$ to $a$. Hence the  Laplace transform of the survival probability $\tilde{S}_a(s|0)= \int_0^\infty S_a(t|0)\, e^{-st}\, dt$ is given by
\begin{equation}
\tilde{S}_a(s|0) = \int_{-\infty}^a e^{-\alpha |x|/(2D)}\, \tilde{\psi}(x,s)\, dx,
\end{equation}
where $\tilde{\psi}(x,s)$ is given by \eref{psi-laplace2-SM}. 
Performing the integral yields
\begin{equation}
\tilde{S}_a(s|0)=\frac{1}{s} \left[1-\frac{p}{A(p)}\, e^{-(\alpha+p) a/(2D)} \right],
\label{S-Laplace}
\end{equation}
with $p=\sqrt{\alpha^2+4Ds}$ and $A(p) = p -\alpha \bigl( 1-e^{-p a/D}\bigr) $. 
The first-passage time distribution is related to the survival probability by 
$F_a(t|x_0)=-\partial_t S_a(t|x_0)$. In the Laplace space, 
this relation reads $\tilde{F}_a(s|x_0)=1-s\tilde{S}_a(s|x_0)$. 
Hence, using \eref{S-Laplace}, we get the Laplace transform of the first-passage distribution as
\begin{equation}
\tilde{F}_a(s|0)=\frac{p}{A(p)}\, e^{-(\alpha+p) a/(2D)}.
\label{F-Laplace}
\end{equation}
    
By inverting formally the Laplace transform \eqref{S-Laplace},
the survival probability can be expressed as a Bromwich integral
\begin{equation}
S_a(t|0)=\frac{1}{2\pi i}\int_{c-i\infty}^{c+i\infty} \tilde{S}(x_0,s) \, e^{st}\, ds.
\label{eq59}
\end{equation}
Let us first consider the trivial limit $a\to\infty$, 
where $\tilde{S}(x_0,s) =1/s$. 
The pole at $s=0$ gives $S_\infty(t|0) =1$, 
which means $\theta(\infty):=-\lim_{t\to \infty} t^{-1} S_\infty (t|0)=0$. 

Now as we bring the absorbing wall from $\infty$ to a finite value $a$, how does the exponent 
$\theta(a):=-\lim_{t\to\infty} t^{-1} S_a(t|0)$ change as a function of $a$? It can be checked from 
\eref{S-Laplace} that, $s=0$ is no longer a pole for any finite values of $a$. Similarly, although $p=0$ 
is a trivial solution of $A(p)=0$, it does not correspond to a pole as it cancels with $p$ in the 
numerator. Nevertheless, $p=0$ is a branch-point.

\begin{figure*}
\includegraphics[width=.45\hsize]{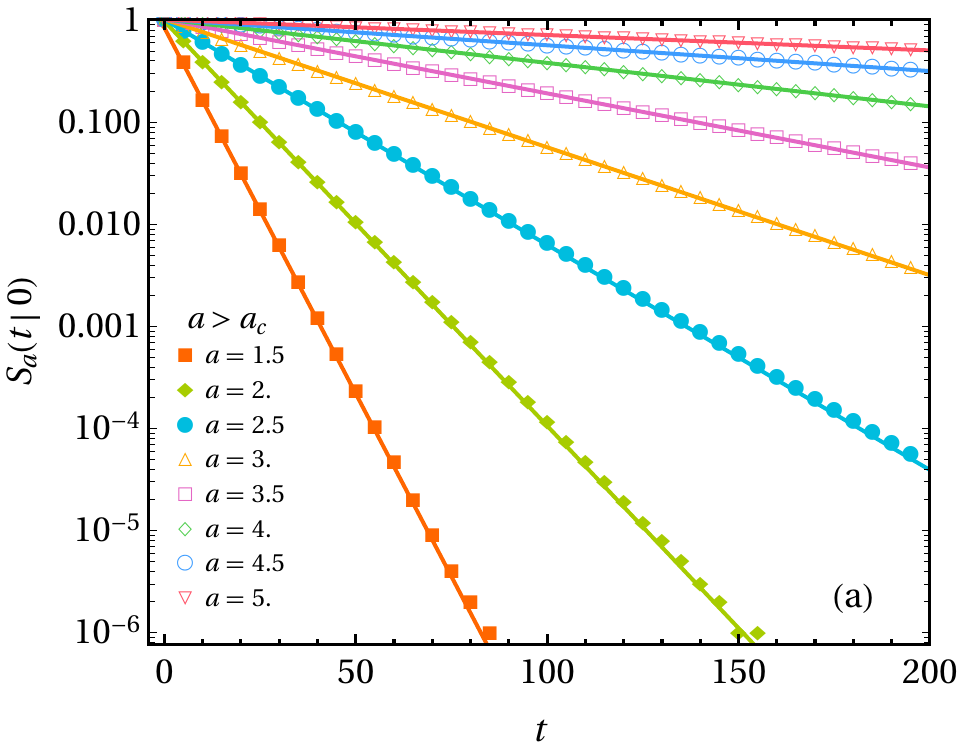}~~
\includegraphics[width=.45\hsize]{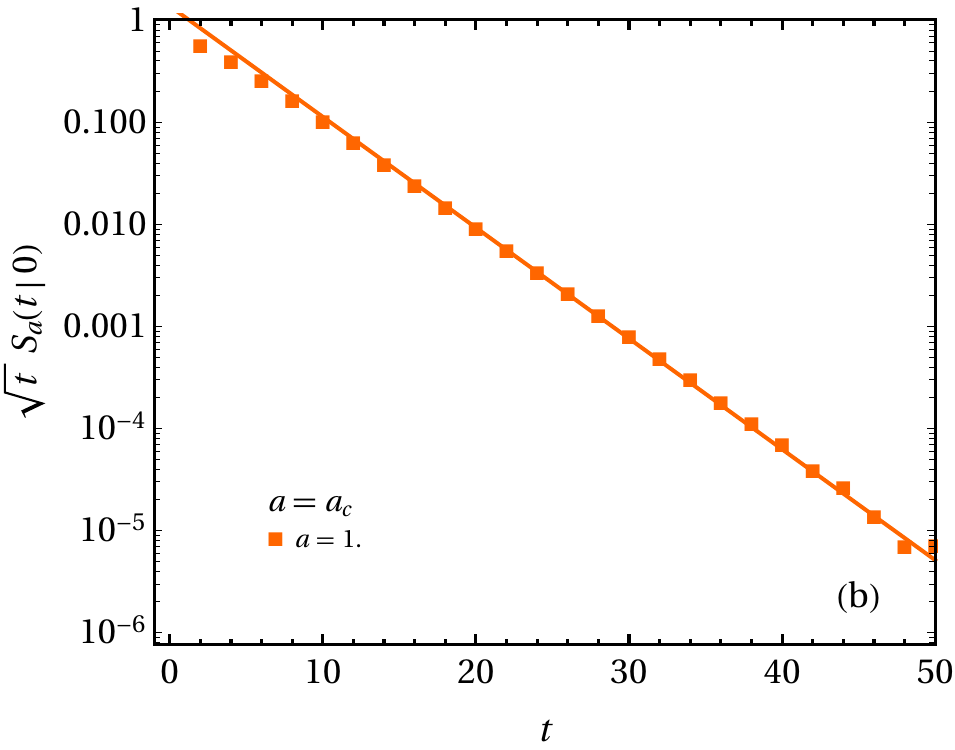}\\[2mm]
\includegraphics[width=.45\hsize]{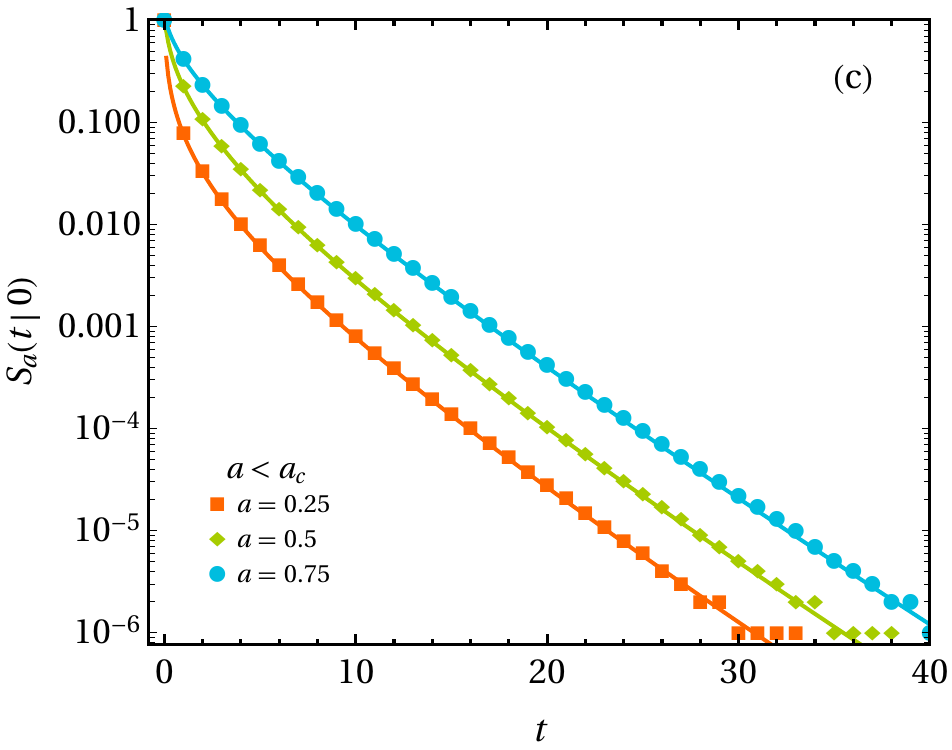}~~
\includegraphics[width=.45\hsize]{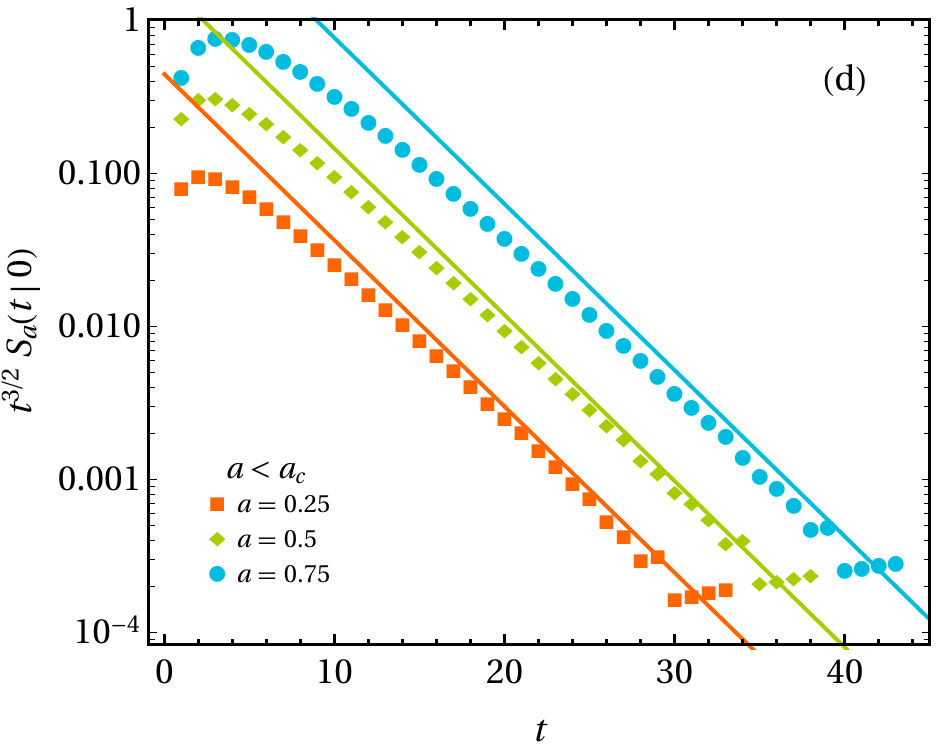}
\caption{\label{surv-fig} Survival probability of a Brownian particle in a potential $U(x)=\alpha |x|$,  starting at the position $x_0=0$   and in the presence of an absorbing barrier at $x=a >0$. We set the diffusion coefficient $D=1$ and $\alpha=1$, so that $a_c=D/\alpha=1$. The points are from numerical simulations whereas the solid lines are theoretical results. In (a) $a>a_c$, the theoretical lines plot the dominant behavior given by the first line of \eref{eq62}. In (b), $a=a_c$, the theoretical line is the leading behavior  from \eref{eq68}. In (c) and (d), $a<a_c$, the theoretical lines plot  \eref{eq69} with $Q(t)$ obtained by numerical integration of \eref{eq64} in (c) and  using \eref{eq66} in (d). 
}

\end{figure*}

For $a > a_c=D/\alpha$, there exists a pole at $s=s^*(a) <0$, on the real-$s$ line, as argued above.
Therefore, for $a>a_c$, the leading contribution at large times, comes from this pole, 
and the branch-point gives the subleading correction:
\begin{align}
S_a(t|0) = R  \, \exp\left(-\frac{\alpha^2-p^{*2}}{4D}\, t\right)   + Q(t) \, \exp\left(-\frac{\alpha^2}{4D}\, t\right) ,
\label{eq62}
\end{align} 
where $R=\lim_{s\to s^*} (s-s^*) \tilde{S}_a(s|0 )$ arises from the residue at $s^*$ and is given by
\begin{align}
R=\frac{2p^{*2} \exp\left[-\frac{a}{2a_c} \,e^{-p^*a/D}\right]}
{\alpha(\alpha+p^{*}) \left[1-\frac{a}{a_c}\,e^{-p^*a/D}\right]}.
\label{eq63}
\end{align}
The prefactor $Q(t)$ of the branch-point contribution arises from the contour  integral around the branch-cut from $-\infty$ to $-\alpha^2/(4D)$, and is given by
\begin{align}
Q(t)= e^{-\alpha a/(2D)}\int_0^\infty dr \,e^{-rt}\,
\frac{\sqrt{4D r}}
     {\pi \bigl[r+\alpha^2/(4D)\bigr]}\,
     \mathrm{Re} \left[-\frac{e^{-ia\sqrt{r/D}}}{A(i\sqrt{4Dr})}\right],
\label{eq64}
\end{align} 
where $\mathrm{Re}[.]$ is the real part of the function inside the square bracket.  
In \fref{surv-fig}~(a), we  compare the leading behavior, i.e., the contribution from the pole, given by the first line of \eref{eq62} with numerical simulation results and find very good agreement.

At $a=a_c$, to the leading order,
$A(p)=p^2/(2\alpha) + O(p^3)$, as $p\to 0$. Therefore, as $a\to a_c^+$, the
pole disappears and there is only a branch-point at $s=-\alpha^2/(4D)$,
corresponding to $p=0$. Note that although the location of the branch-point
is independent of $a$, the nature of the singularity  for $a=a_c$ is
different from that for $a\not=a_c$. At, $a=a_c$, following
Eqs.~\eqref{eq64}, the leading order behavior for large  $t$ is given by
\begin{equation}
S_{a_c}(t|0)= Q(t)\,  \exp\left(-\frac{\alpha^2}{4D}\, t\right),
\label{eq68}
\end{equation}
where
\begin{equation}
Q(t)= 
e^{-\frac{1}{2}}\,
\frac{4\sqrt{D}}{\alpha\sqrt{\pi\, t}}   +O(t^{-3/2}),
\label{eq68.1}
\end{equation}
\Fref{surv-fig}~(b)  compares this leading behavior with numerical simulation and shows a very good agreement.  

For $a<a_c$, there is no pole [see \fref{p-eqn2-fig}], and only a branch-point at $s=-\alpha^2/(4D)$. Therefore, 
\begin{equation}
S_a(t|0) =  Q(t) \, \exp\left(-\frac{\alpha^2}{4D}\, t\right),
\label{eq69}
\end{equation} 
where $Q(t)$ is given by \eref{eq64}. 
For  $ a \ll a_c$, the leading order behavior for large $t$  can be evaluated as 
\begin{equation}
Q(t)= \frac{e^{-\alpha a/(2D)}} {(1-a/a_c)^2} 
\frac{2a\sqrt{D} }{\alpha^2\sqrt{\pi}\, t^{3/2}} 
 +  O(t^{-5/2}),
 \label{eq66}
\end{equation}

Figures~\ref{surv-fig}~(c) and (d)  compare this result with numerical simulations, where $Q(t)$ evaluated exactly by performing numerical integration of  \eref{eq64} in (c), and the leading order behavior of $Q(t)$ given by \eref{eq66} is used in (d). Naturally, the agreement is perfect in \fref{surv-fig}~(c) whereas in \fref{surv-fig}~(d), it becomes better as $a$ moves away from $a_c$. 

To summarize, the leading asymptotic of $S_a(t|0)$ is given by
\begin{equation}
S_a(t|0) \sim \exp\bigl[-\theta(a) t\bigr],
\label{eq70}
\end{equation}
where 
\begin{equation}
\theta(a) = \begin{cases}\displaystyle
\frac{1}{4D} 
\bigl[\alpha^2- p^{*2}(a)\bigr] &\quad\text{for}~ a>a_c\\[3mm]
\displaystyle
\frac{\alpha^2}{4D} 
 &\quad\text{for}~ a<a_c
\end{cases}
\label{eq71}
\end{equation}
in which $0<p^*(a)<\alpha $ solves the   transcendental
equation  $p =\alpha \bigl( 1-e^{-p a/D}\bigr)$ [see \fref{p-eqn2-fig}]. Note that while $\theta(a)$ varies as a function of $a$ for $a>a_c$, it remains at the constant value $\alpha^2/(4D)$ for $a<a_c$. Near $a=a_c^+$,   we have $p^* = 2D (a-a_c)/a_c^2 + O\bigl[(a-a_c)^2\bigr]$.
Therefore, as $a$ approaches $a_c$ from above, 
\begin{equation}
\theta(a) = \frac{\alpha^2}{4D} - \frac{D}{a_c^4}(a-a_c)^2 +O\bigl[ (a-a_c)^3\bigr].
\end{equation}

\vspace{0.3cm}

\begin{figure}
\includegraphics[width=.5  \hsize]{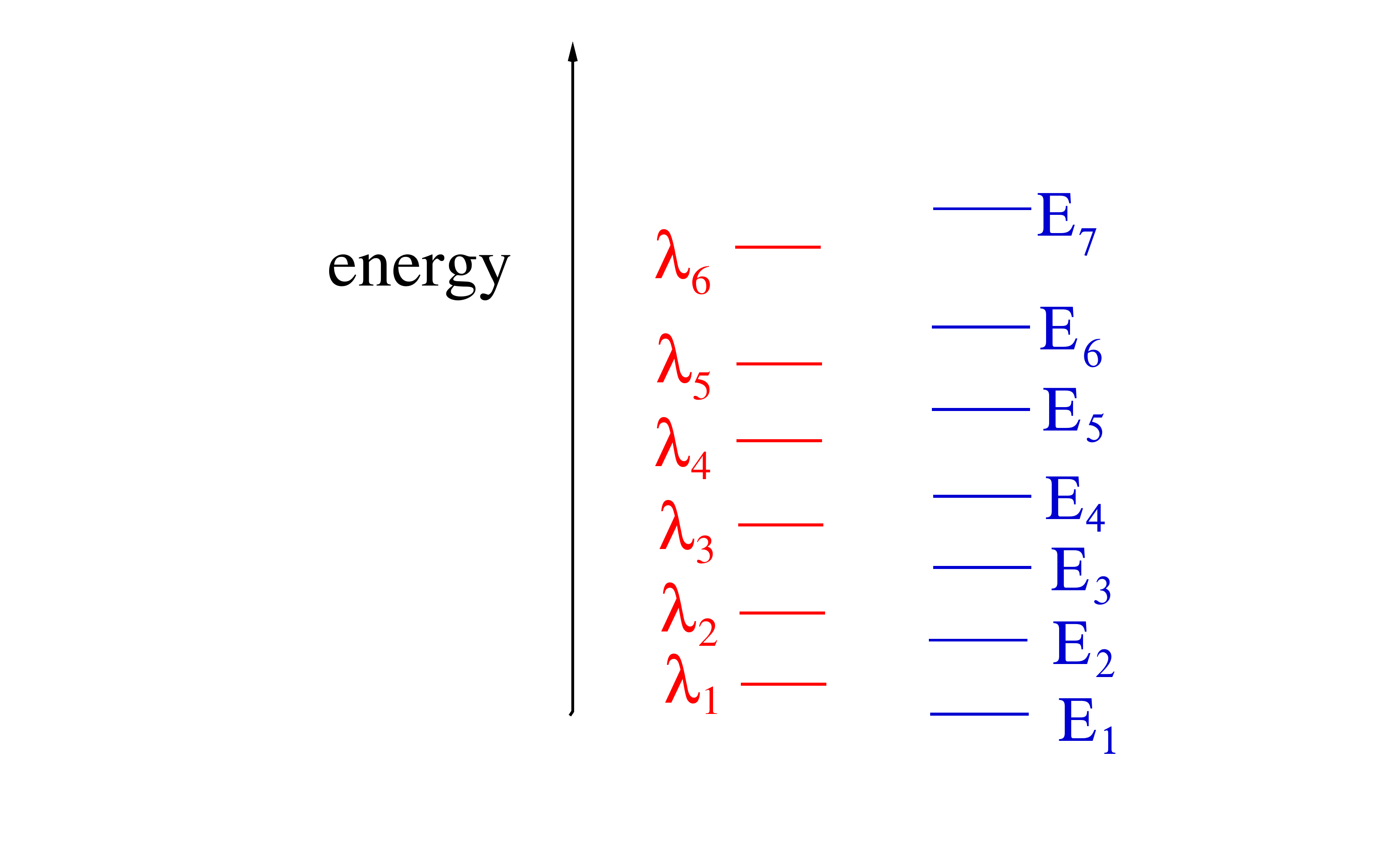}
\caption{For a sufficiently confining potential in the 
unconstrained system (with the absorbing barrier $a\to \infty$),
the discrete energy eigenvalues $\{E_1<E_2<E_3\ldots\}$
of the associated Schr\"odinger equation are shown schematically
on the right (blue). As the barrier location $a\ge 0$ decreases,
the new ordered eigenvalues (first-passage eigenvalues)
$\{\lambda_1(a)<\lambda_2(a)<\lambda_3(a)\ldots\}$ are
shown (schematic) on the left (red). The sets $\{E_i\}$ and
$\{\lambda_i(a)\}$ interlace, i.e., between any two consecutive
$E_i$'s,  there is one and only one $\lambda_i$, as shown schematically
in the figure.} 
\label{spectrum_fig}
\end{figure}

{\noindent {\bf {Link to the interlace theorem:}}} We now show
how our results are consistent with a powerful
interlace theorem derived in Refs.~\cite{HG2018,HG2019}.
Let us first consider an unconstrained motion
of particle (no absorbing barrier or equivalently $a\to \infty$ limit) 
diffusing in a sufficiently confining potential $U(x)$, such that its
quantum counterpart $V(x)=[U'(x)]^2/{4D}- U''(x)/2$ has a discrete spectrum say 
$\{E_1<E_2< E_3< \ldots\}$. These energy eigenvalues of the Schr\"odinger
equation in the quantum potential $V(x)$ are identical to
the relaxation eigenvalues of the unconstrained Fokker-Planck
equation, $\partial_t P- D\partial_x^2 P- \partial_x [U'(x)P]=0$ on
the full real line.
Now, imagine that we bring the absorbing barrier from $a=\infty$ to
a finite value $a\ge 0$, i.e., the barrier always stays
to the right of the initial position of the particle which we set to be $x=0$. 
If the barrier is to the left of the particle, one can reverse $x$ and
it would be a similar analysis. In this paper, we therefore restrict ourselves to $a\ge 0$.
The presence of the absorbing barrier at $a\ge 0$
changes the spectrum of the Fokker-Planck operator. The new ordered eigenvalues 
$\{\lambda_1(a)<\lambda_2(a)<\lambda_3(a) \ldots\}$ obviously
depend on $a$. Note that the lowest eigenvalue $\lambda_1(a)\equiv \theta(a)$
by definition.
Hence as $a$ decreases towards $0$, the whole spectrum $\{\lambda_i(a)\}$ 
moves upwards. The unconstrained spectrum $\{E_i\}$ of course is independent
of $a$.
According to the interlace theorem, for any given $a$, the $\{\lambda_i\}$'s
are interlaced with $\{E_i\}$'s, i.e., $E_1 \le \lambda_1(a) \le E_2 \le \lambda_2(a) \le E_3 \le \dotsb$,  ---between any pair
of consecutive $E_i$'s, there is  one and only one $\lambda_i$ (see
Fig. (\ref{spectrum_fig})). As $a$
decreases, the whole spectrum $\{\lambda_i(a)\}$ slides upwards, but
always respects the interlacing rule for any $a\ge 0$. In this case, the lowest eigenvalue
$\lambda_1(a)=\theta(a)$ monotonically increases with decreasing $a$ as $a\to 0$.
As mentioned above, this occurs for sufficiently confining potentials, e.g.,
when $U(x)\sim |x|^{\gamma}$ as $x\to -\infty$ with $\gamma>1$.

In the case $\gamma=1$, e.g., for $U(x)= \alpha |x|$ as $x\to -\infty$,
the Schr\"odinger equation in
the full space (i.e., when the barrier location $a\to \infty$) has
a single discrete eigenvalue (the lowest one) and 
a band of continuous eigenvalues, separated from the ground state
by a finite gap $\alpha^2/(4D)$, as shown in Fig. 2 in the main text. As $a$
decreases, the continuous part of the spectrum does not change,
but the lowest discrete eigenvalue increases (see Fig. 2
in the main text). Since the continuous part
of the spectrum is independent of $a$, clearly the discrete eigenvalue can not
increase beyond the gap $\alpha^2/(4D)$. Had it done so, 
it will go over the lowest eigenvalue of the continuous sector 
and clearly violate the interlacing rule mentioned above. Hence
$\theta(a)$ freezes when it hits the gap value $\alpha^2/(4D)$, at
a certain critical value $a_c=D/\alpha$ as shown in our paper.
Thus this freezing transition and the mechanism behind it is completely
consistent with the interlacing theorem, applied to the discrete
part of the spectrum in the critical case $\gamma=1$ till the
onset of the freezing transition.

\section{The backward Fokker-Planck equation for the survival probability in the Laplace space}

As mentioned in the main text, the backward Fokker-Planck equation satisfied by the survival probability, with the starting position $x_0$, reads
\begin{equation}
\frac{\partial S_a}{\partial t} = D \frac{\partial^2 S_a}{\partial x_0^2} -  U'(x_0)\, \frac{\partial S_a}{\partial x_0}
\end{equation} 
with the initial condition $S_a(0|x_0)=1$ and the boundary conditions, 
$S_a(t|x_0\to -\infty)=1$ and $S_a(t|x_0=a)=0$. 
The Laplace transform  $\tilde{S}_a(s|x_0)= \int_0^\infty S_a(t|x_0)\, e^{-st}\, dt$ then satisfies
\begin{equation}
D \tilde{S}_a''(s|x_0) -U'(x_0) \tilde{S}_a'(s|x_0) -s \tilde{S}_a(s|x_0) = -1,
\label{S-Laplace.1}
\end{equation}
with the boundary conditions
\begin{equation}
\tilde{S}_a(s|x_0\to-\infty)=1/s\quad\text{and}\quad \tilde{S}_a(s|x_0=a)=0.
\end{equation}
Setting
\begin{equation}
\tilde{S}_a(s|x_0) =  \frac{1}{s} + \tilde{q}(x_0),
\end{equation}
$\tilde{q}(x_0)$ satisfies the homogeneous differential equation
\begin{equation}
D \tilde{q}''(x_0) -U'(x_0) \tilde{q}'(x_0) -s \tilde{q}(x_0) = 0,
\label{q2}
\end{equation}
with the boundary conditions $\tilde{q}(x_0\to -\infty)=0$ and $\tilde{q}(x_0=a)=-1/s$.

\section{The details of computing the survival probability for a 
potential that  grows faster than $|x|$ as\,  $x\to -\infty$}

The example of the potential considered in the main text is 
\begin{equation}
U(x) = \begin{cases}
 \frac{1}{2}\, \mu \, x^2&\text{for}~ x <-b, \\
 \alpha |x| &\text{for}~ -b <x < a,
\end{cases}
\label{potential (i)}
\end{equation}
where $b >0$ and for the sake of continuity of the potential at $x=-b$, we set $\mu =2\alpha/b$.

We have to solve \eref{q2} separately in the three regions, (i)  $x_0<-b$, (ii) $-b < x_0 <0$,  and (iii) $0<x_0 <a$ and then match the solutions for the continuity of the solutions as well as the derivatives at both $x_0=-b$ and $x_0=0$. For $x_0 <-b$, \eref{q2} reads
\begin{equation}
D\tilde{q}''(x_0) -\mu x_0 \tilde{q}'(x_0) -s\tilde{q}(x_0)=0.
\label{qx0}
\end{equation}
Now substituting 
\begin{equation}
\tilde{q}(x_0)=e^{\mu x_0^2/(4 D)}\, w(x_0\sqrt{\mu/D}), 
\end{equation}
above gives the differential equation
\begin{equation}
w''\bigl(x_0\sqrt{\mu/D}\bigr)  + \left[-\frac{s}{\mu} + \frac{1}{2} - \frac{1}{4} \bigl(x_0\sqrt{\mu/D}\bigr)^2 \right] w\bigl(x_0\sqrt{\mu/D}\bigr) =0, 
\end{equation}
whose general solution can be expressed in terms of the two linearly independent parabolic cylinder functions  $D_{-s/\mu}\bigl(x_0\sqrt{\mu/D}\bigr)$ and $D_{-s/\mu} \bigl(-x_0\sqrt{\mu/D}\bigr)$.  Noting that $D_{-s/\mu}\bigl(x_0\sqrt{\mu/D}\bigr)\to \infty$  whereas $D_{-s/\mu}\bigl(-x_0\sqrt{\mu/D}\bigr)\to 0$ as $x_0 \to -\infty$, the solution of \eref{qx0} that tends to zero as $x_0\to -\infty$, is given by
\begin{equation}
\tilde{q}(x_0)= A_1 \, e^{\mu  x_0^2/(4 D)} D_{-\frac{s}{\mu}}\left(-x_0 \sqrt{\frac{\mu}{D}}\right),
\end{equation}
 where  the constant $A_1$ to be determined.  

For $-b < x_0 < 0$, \eref{q2} reads
\begin{equation}
D \tilde{q}''(x_0) +\alpha  \tilde{q}'(x_0) -s \tilde{q}(x_0) = 0.
\end{equation} 
The general solution is given by
\begin{equation}
\tilde{q}(x_0)= A_2\, e^{(p-\alpha) x_0/(2D)} +  B_2  e^{-(p+\alpha) x_0/(2D)},
\end{equation}
with $p=\sqrt{\alpha^2+4Ds}$ and the constants $A_2$ and $B_2$ to be determined.

Finally, for $0 <x_0 <a$, \eref{q2} reads
\begin{equation}
D \tilde{q}''(x_0) -\alpha  \tilde{q}'(x_0) -s \tilde{q}(x_0) = 0,
\end{equation} 
whose general solution is given by
\begin{equation}
\tilde{q}(x_0)= A_3\, e^{(\alpha+p) x_0/(2D)} +  B_3  e^{(\alpha-p) x_0/(2D)}
\end{equation}
with the constant $A_3$ and $B_3$ to be determined.

After determining the five constants by using the four matching conditions at $x_0=-b$ and $x_0=0$,  and the boundary condition at $x_0=a$,  for the starting position $x_0=0$ (taking for simplicity), we get 
\begin{equation}
\tilde{S}_a (s|0) = \frac{1}{s} \bigl[1-\tilde{F}_a(s|0)\bigr]
\end{equation}
where the Laplace transform of the first-passage time  distribution is given by
\begin{equation}
\tilde{F}_a(s|0)=\frac{p}{B(s)}\, e^{-(\alpha+p) a/(2D)}\, \chi(s),
\end{equation}
with $p=\sqrt{\alpha^2+4 D s}$ and 
\begin{align}
B(s) &= \left[p^2 \left(e^{\frac{p b}{D}}+e^{-\frac{p a}{D}}\right)+3 \alpha ^2 \left(1-e^{-\frac{ p a}{D}}\right) \left(e^{\frac{p b}{D}}-1\right)-\alpha  p \left(4 e^{\frac{p b}{D}}+1-e^{\frac{ p b}{D}} e^{-\frac{p a}{D}}-4 e^{-\frac{p a}{D}}\right)\right] D_{-\frac{s}{\mu }}\left(b \sqrt{\frac{\mu }{D}}\right)\notag\\
&\quad-2 D \left[\alpha  \left(1-e^{-\frac{a p}{D}}\right) \left(e^{\frac{b p}{D}}-1\right)+p \left(e^{-\frac{pa}{D}}-e^{\frac{p b}{D}}\right)\right]\, \sqrt{\frac{\mu }{D}} \,  D_{-\frac{s}{\mu }+1}\left(b \sqrt{\frac{\mu }{D}}\right),
\\
\chi(s)& =\left[p+3 \alpha +(p-3 \alpha ) e^{\frac{b p}{D}}\right] D_{-\frac{s}{\mu }}\left(b \sqrt{\frac{\mu }{D}}\right)+2 D \left(e^{\frac{p b}{D}}-1\right) \, \sqrt{\frac{\mu }{D}} \,\, D_{-\frac{s}{\mu }+1}\left(b \sqrt{\frac{\mu }{D}}\right)
.    
\end{align}

After substituting, $s=(p^2-\alpha^2)/(4D)$, it is easy to check that $\tilde{F}_a(s(p)|0)\equiv \tilde{G}_a(p)$ is a symmetric function with respect to $p$, indicating that series expansion of $\tilde{G}_a(p)$ around $p=0$ contains only even powers of $p$.  Therefore, $p=0$, and equivalently $s=-\alpha^2/(4D)$, is not a branch-point. Similarly, it is easy to check that $s=-\alpha^2/(4D)$ is not a pole of $\tilde{F}_a(s|0)$, i.e., $p=0$ is not a pole of $\tilde{G}_a(p)$.

\begin{figure}
\includegraphics[width=.5  \hsize]{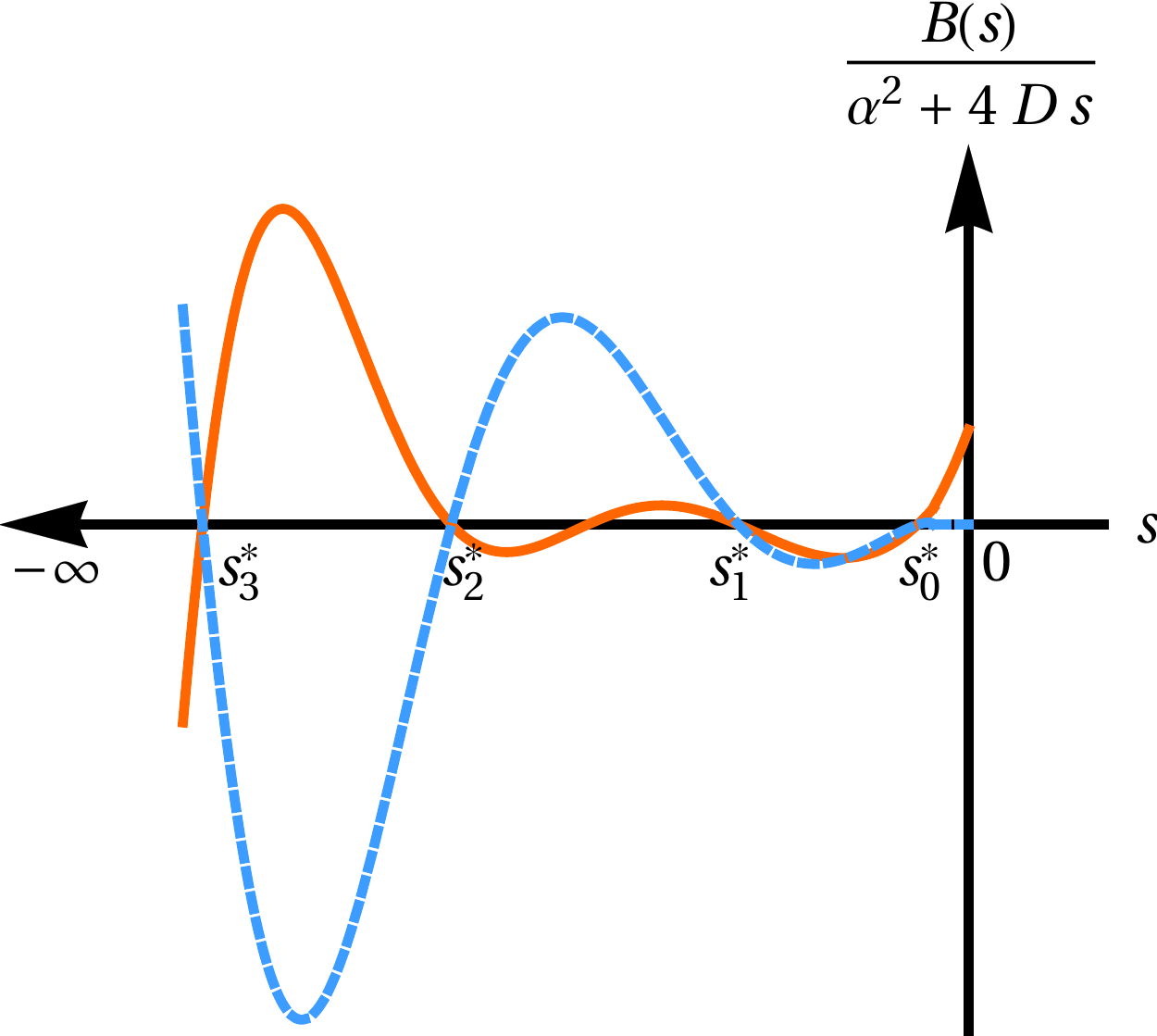}
\caption{$B(s)/(\alpha^2+4Ds)$ as a 
function $s$ for $s \le 0$, for certain parameter values $\alpha=D=\mu=1$ and $b=2$. 
The red solid line and the blue dashed line plot respectively the real part and the imaginary part 
of the function. The zeros of the function given by the points where both real and 
imaginary part become zero.  }
\label{bs}
\end{figure}

\begin{figure}
\includegraphics[width=.49\hsize]{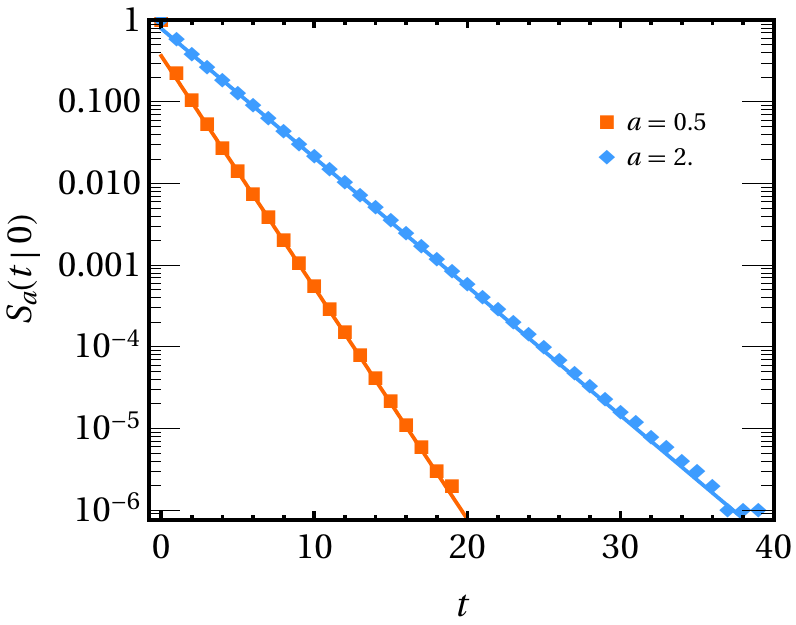}~~\includegraphics[width=.49\hsize]{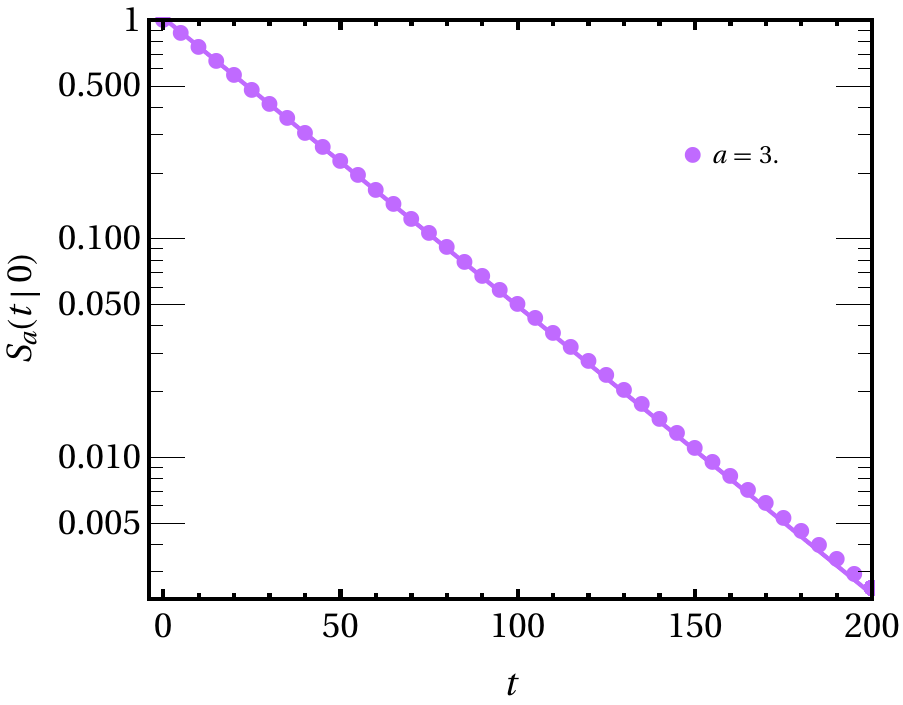}~
\caption{\label{s-fig-quadratic}Survival probability of a Brownian particle in a potential given by \eref{potential (i)},  starting at the position $x_0=0$  and in the presence of an absorbing barrier at $x=a >0$, for $a=0.5$ and $2.0$ on the left panel and $a=3.0$ om the right panel. We set $D=\alpha=\mu=1$ and $b=2$. The points are from numerical simulations whereas the solid lines plot the leading order theoretical expression given by \eref{s-quadratic-asym}.   }
\end{figure}

Although $s=-\alpha^2/(4D)$ is a zero of $B(s)$, it cancels with the numerator as both the numerator and the denominator tend to zero as $\alpha^2+4 Ds$ near $s=-\alpha^2/(4D)$. Therefore, to find the poles of $\tilde{S}_a(s|0)$ one should look at the zeros of $B(s)/(\alpha^2+4Ds)$. This function has infinite number of zeros on the negative $s$ axis [see~\fref{bs}], denoted by $-\infty <\dotsb < s_2^*(a) < s_1^*(a)  < s_0^*(a)  <0$. 
Therefore, inverting the Laplace transform, the survival probability is given by 
\begin{equation}
S_a(t|0) = \sum_{i=0}^\infty\, R_i (a) \,  e^{s_i^*(a) t}, 
\end{equation}
where 
\begin{equation}
R_i(a) = \lim_{s\to s_i^*} (s-s_i^*)\, \tilde{S}_a(s|0) = \frac{\sqrt{\alpha^2+4 D s_i^*(a)}\,\,  \chi(s_i^*(a))}{\bigl[-s_i^*(a)\bigr]\,\,  B'(s_i^*(a))}\, \exp\left(-\frac{a}{2 D}\left[\alpha + \sqrt{\alpha^2 + 4 D s_i^*(a)}\right]\right).
\end{equation}
Evidently, the leading behavior of the survival probability at large times is given by
\begin{equation}
S_a(t|0) = R_0(a)\,  e^{-\theta(a) \, t} + O(e^{s_1^*(a) t}), 
\label{s-quadratic-asym}
\end{equation}
where $\theta(a)=-s_0^*(a)$. We compare this expression with numerical simulation in \fref{s-fig-quadratic} and find very good agreement. 

\section{ The details of computing the survival probability for a potential that grows slower than $|x|$ as $x\to -\infty$} 
\label{logx-potential-left}

The example of the potential considered in the main text is 
\begin{equation}
U(x) = \begin{cases}
 c \ln (-x/\lambda) &\text{for}~ x <-b, \\
 \alpha |x| &\text{for}~ -b <x < a,
\end{cases}
\label{potential (iii)}
\end{equation}
where $\lambda>0$ sets a length scale, $b >0$,  and for the sake of continuity of the potential at $x=-b$, we set $c\ln (b/\lambda)=\alpha b$.

We have to solve \eref{q2} separately in the three regions, (i)  $x_0<-b$, (ii) $-b < x_0 <0$,  and (iii) $0<x_0 <a$ and then match the solutions for the continuity of the solutions as well as the derivatives at both $x_0=-b$ and $x_0=0$. In the region $x<-b$, \eref{q2} reads
\begin{equation}
D \tilde{q}''(x_0) -(c/x_0)\,  \tilde{q}'(x_0) -s \tilde{q}(x_0) = 0.
\end{equation} 
The solution of this equation, that tends to zero as $x\to -\infty$, is given by
\begin{equation}
\tilde{q}(x_0)=A_1 |x_0|^\nu K_\nu \bigl(|x_0|\sqrt{s/D}\bigr),
\end{equation}
where $A_1$ is a constant to be determined, $\nu=(c/D + 1)/2$,  and $K_\nu(z)$ is the modified Bessel function of the second kind.

In the region $-b <x_0 <0$, \eref{q2} reads
\begin{equation}
D \tilde{q}''(x_0) +\alpha  \tilde{q}'(x_0) -s \tilde{q}(x_0) = 0.
\end{equation} 
The general solution is given by
\begin{equation}
\tilde{q}(x_0)= A_2\, e^{(p-\alpha) x_0/(2D)} +  B_2  e^{-(p+\alpha) x_0/(2D)},
\end{equation}
with $p=\sqrt{\alpha^2+4Ds}$ and the constants $A_2$ and $B_2$ to be determined.

Finally, for $0 <x_0 <a$, \eref{q2} reads
\begin{equation}
D \tilde{q}''(x_0) -\alpha  \tilde{q}'(x_0) -s \tilde{q}(x_0) = 0,
\end{equation} 
whose general solution is given by
\begin{equation}
\tilde{q}(x_0)= A_3\, e^{(\alpha+p) x_0/(2D)} +  B_3  e^{(\alpha-p) x_0/(2D)}
\end{equation}
with the constant $A_3$ and $B_3$ to be determined.

After determining the five constants by using the four matching conditions at $x_0=-b$ and $x_0=0$,  and the boundary condition at $x_0=a$,  for the starting position $x_0=0$ (taking for simplicity), we get  
\begin{equation}
\tilde{S}_a (s|0) = \frac{1}{s} \bigl[1-\tilde{F}_a(s|0)\bigr]
\end{equation}
where the Laplace transform of the first-passage time  distribution is given by
\begin{equation}
\tilde{F}_a(s|0)=\frac{p}{B(s)}\, e^{-(\alpha+p) a/(2D)}\, \chi(s),
\end{equation}
with $p=\sqrt{\alpha^2+4 D s}$ and 
\begin{align}
B(s) &= 
\left[p^2 e^{p b/D}-\left(\alpha-(\alpha +p) e^{-p a/D}\right) \left(\alpha  
\left(e^{p b/D}-1\right)+p\right)\right] K_{\nu }\left(b \sqrt{\frac{s}{D}}\right) \notag\\
&\quad+2 D  \left[p \left(e^{p b/D}-e^{-pa/D}\right)-\alpha  \left(1-e^{-p a/D}\right) \left(e^{p b/D}-1\right)\right] \sqrt{\frac{s}{D}}\,K_{\nu -1}\left(b \sqrt{\frac{s}{D}}\right),
\\
\chi(s)& =
\left[p-\alpha +(\alpha +p) e^{p b/D}\right] K_{\nu }\left(b \sqrt{\frac{s}{D}}\right)
+2 D  \left(e^{p b/D}-1\right) \, \sqrt{\frac{s}{D}}\, K_{\nu -1}\left(b \sqrt{\frac{s}{D}}\right).    
\end{align}

Note that  in the limit $a\to\infty$, we get $\tilde{F}_a(s|0)\to 0$,  and consequently, $\tilde{S}_a(s|0) \to 1/s$. This gives, $\lim_{a\to\infty} S(t|0) =1$ at all times,  as expected. 
For any finite $a$, 
after substituting, $s=(p^2-\alpha^2)/(4D)$, it is easy to check that $\tilde{F}_a(s|0)$ is a symmetric function with respect to $p$, indicating that series expansion of $\tilde{F}_a$ around $p=0$ contains only even powers of $p$.  Therefore, $p=0$, and equivalently $s=-\alpha^2/(4D)$, is not a branch-point. 
For several sets of  representative values of the parameters, we have numerically checked that $\tilde{F}_a(s|0)$ does not have any pole, indicating a non-exponential late time behaviour for $F_a(t|0)$ as well as for $S_a(t|0)$.  Anticipating such non-exponential behaviour to show up as branch-point singularities, we analyse $\tilde{F}_a(s|0)$ near $s=0$.  In our case, we choose the potential $U(x)$ to be confining so that there is an equilibrium state (for $a\to \infty$) given by the Boltzmann distribution $P(x, t\to \infty) \propto e^{-U(x)/D}$ that is integrable, i.e.,  $\int e^{-U(x)/D}\, dx$ is finite. This requires $c/D > 1$, and consequently, $\nu = (1+c/D)/2 >1$.

Using the leading behaviour of the modified Bessel function of the second kind near $s=0$, for  a given non-integer $\nu  \in (n,n+1)$ with $n\ge 1$ being an integer,
we get
\begin{align}
K_{\nu }\left(b \sqrt{\frac{s}{D}}\right) &= s^{-\nu/2} \left[ \sum_{m=0}^n \alpha_m \, s^m + \alpha_\nu\, s^\nu + O\bigl(s^{n+1}\bigr)\right],\\
\sqrt{\frac{s}{D}}\, K_{\nu -1}\left(b \sqrt{\frac{s}{D}}\right) &=
s^{-\nu/2} \left[ \sum_{m=1}^n \beta_m \, s^m + \beta_\nu\, s^\nu + O\bigl(s^{n+1}\bigr)\right],
\end{align}
where the coefficients in the above series can be obtained explicitly. 
In particular, 
\begin{align}
\alpha_0&=2^{\nu -1} b^{-\nu } D^{\nu /2} \Gamma (\nu ),
 \quad 
\alpha_\nu = 2^{-\nu -1} b^{\nu } D^{-\frac{\nu }{2}} \Gamma (-\nu ) , \\
\beta_1&= 2^{\nu -2} b^{1-\nu } D^{\frac{\nu }{2}-1} \Gamma (\nu -1), \quad\text{and} \quad
\beta_\nu = 2^{-\nu } b^{\nu -1} D^{-\frac{\nu }{2}} \Gamma (1-\nu ).
\end{align}

The remaining factors in both $B(s)$ and $\chi(s)$ are analytic about $s=0$. Therefore, for a given (finite and fixed)  value of $a$, $\tilde{F}_a(s|0)$ can be expressed as
\begin{equation}
\tilde{F}_a(s|0) = \frac{\bigl[\sum_{j=0}^\infty a_j s^j\bigr]  \left[ \sum_{m=0}^n \alpha_m \, s^m + \alpha_\nu\, s^\nu + O\bigl(s^{n+1}\bigr)\right] + \bigl[ \sum_{j=0}^\infty b_j s^j\bigr]\left[ \sum_{m=1}^n \beta_m \, s^m + \beta_\nu\, s^\nu + O\bigl(s^{n+1}\bigr)\right] }{\bigl[\sum_{j=0}^\infty c_j s^j\bigr]  \left[ \sum_{m=0}^n \alpha_m \, s^m + \alpha_\nu\, s^\nu + O\bigl(s^{n+1}\bigr)\right] + \bigl[ \sum_{j=0}^\infty d_j s^j\bigr]\left[ \sum_{m=1}^n \beta_m \, s^m + \beta_\nu\, s^\nu + O\bigl(s^{n+1}\bigr)\right] }, 
\label{fp-series1}
\end{equation}
where the coefficients $\{ a_j, b_j, c_j, d_j\}$ can be computed. In particular, the first few coefficients are given by
\begin{align}
a_0&= c_0=2 \alpha ^2 e^{-\frac{\alpha  a}{D}} e^{\frac{\alpha  b}{D}},\quad
a_1= 2 e^{-\frac{a \alpha }{D}} \left[e^{\frac{\alpha  b}{D}} (-a \alpha +2 \alpha  b+3 D)+D\right],
\quad b_0=2 \alpha  D e^{-\frac{a \alpha }{D}} \left(e^{\frac{\alpha  b}{D}}-1\right), \\
c_1&=
e^{-\frac{a \alpha }{D}} \left[2 D \left(2 e^{\frac{\alpha  (a+b)}{D}}-e^{\frac{a \alpha }{D}}+e^{\frac{\alpha  b}{D}}+2\right)-4 \alpha  (a-b) e^{\frac{\alpha  b}{D}}\right], \quad
d_0=2 \alpha  D e^{-\frac{a \alpha }{D}} \left(e^{\frac{a \alpha }{D}}+e^{\frac{\alpha  b}{D}}-2\right).
\end{align}
\Eref{fp-series1} can be expanded in a series as 
\begin{equation}
\tilde{F}_a(s|0)=\sum_{m=0}^n \frac{(-1)^m}{m!}\, T_{a,m}(0)\, s^m + \nu\, \Gamma(-\nu)\, a_\nu\,  s^{\nu} +o(s^\nu),
\end{equation} 
where $T_{a,m}(0)=\int_0^\infty t^m\, F_a(t|0)\, dt$ is the $m$-th moment of the first-passage time distribution. 
In particular, we have $T_{a,0}(0)=a_0/c_0=1$, as expected from normalization of the first-passage time distribution.  The mean first-passage time is given by
\begin{equation}
T_a(0)\equiv T_{a,1}(0)  = \frac{1}{a_0\alpha_0} \left[ \alpha_0 (c_1-a_1) + \beta_1 (d_0-b_0)\right],
\end{equation}
which, after explicit evaluation, agrees with  \eref{Ta-3} obtained from  the general expression 
 \eref{mean fp}.

The coefficient of the 
most dominant non-analytic term is given by
\begin{equation}
\nu\,\Gamma(-\nu)\, a_\nu = \frac{\beta_\nu (b_0-d_0)}{a_0\alpha_0} ,
\end{equation}
which after evaluation (while using $c\ln (b/\lambda) = \alpha b$) gives 
\begin{equation}
a_\nu = 
\left(\frac{2D}{\alpha \lambda}\right)\,  \left(e^{\frac{\alpha a}{D}}-1\right)\, \frac{1}{\Gamma(\nu)}\, \left(\frac{\lambda^2}{4D} \right)^\nu\, .
\end{equation}
 Using the relation 
\begin{equation}
\int_0^\infty\, dz \left[e^{-z} - \sum_{m=0}^n \frac{(-z)^m}{m!} \right]\, {z^{-(\nu+1)}} = \Gamma(-\nu)
\quad \text{for $n < \nu < n+1$,}
\end{equation}
  we find that the late time behavior of the first-passage time distribution is given by the power-law distribution 
\begin{equation}
F_a(t|0)  = \nu\,  a_\nu\, t^{-(\nu+1)} + o( t^{-(\nu+1)}).
\end{equation}
Consequently, the survival probability has the asymptotic power-law decay
\begin{equation}
S_a(t|0) = \int_t^\infty F_a(t'|0)\, dt' =  a_\nu\, t^{-\nu} + o(t^{-\nu}). 
\label{S-long-t-log}
\end{equation}
For integer values of $\nu=n$, the Bessel's functions have $\ln s$ singularities, which in turn 
gives rise to $\ln s$ singularities in $\tilde{F}_a(s|0)$. Consequently,  the first-passage time distribution and survival probability 
have power-low decays accompanied by $\ln t$ corrections.

\section{The mean first-passage time}

The mean first-passage time $T_a(x_0)$ to a position $a$, starting with the position $x_0 <a $ 
can be computed exactly for arbitrary confining potential $U(x)$~\cite{Risken1984}. Here, for convenience,
we reproduce this proof. The mean first-passage time
is defined by
\begin{equation}
T_a(x_0)= \int_0^\infty\, t \, F_a (t|x_0)\, dt.
\end{equation}
Using $F_a(t|x_0)=-\partial_t S_a(t|x_0)$ in the above integral and then integrating  by parts using the boundary conditions $S_a(t=0|x_0)=1$ and $S_a(t\to\infty|x_0)=0$, we get
\begin{equation}
T_a(x_0)=\int_0^\infty S_a(t|x_0)\, dt \equiv \tilde{S}_a(s=0|x_0).
\end{equation}
Therefore, setting $s=0$ in  \eref{S-Laplace}, we get the differential equation
\begin{equation}
D \frac{d^2 T_a}{d x_0^2} - U'(x_0) \frac{dT_a}{dx_0} =-1.
\end{equation} 
Using $W(x_0) = T_a'(x_0)$, we get
\begin{equation}
\frac{dW}{dx_0} - \frac{U'(x_0)}{D} W(x_0) = -\frac{1}{D}.
\end{equation}
Multiplying both sides of  the above equation by $e^{-U(x_0)/D}$ we get 
\begin{equation}
\frac{d}{dx_0} \left[ W(x_0)\, e^{-U(x_0)/D}\right] = -\frac{1}{D}\, e^{-U(x_0)/D},
\end{equation}
which can be integrated to 
\begin{equation}
W(x_0) e^{-U(x_0)/D} = -\frac{1}{D} \int_{-\infty}^{x_0} e^{-U(z)/D}\, dz,
\end{equation}
where we have used the boundary condition $\lim_{x_0\to -\infty} W(x_0) e^{-U(x_0)/D} = 0$. Therefore, 
\begin{equation}
\frac{d T_a}{d x_0} = -\frac{1}{D} e^{U(x_0)/D}\,  \int_{-\infty}^{x_0} e^{-U(z)/D}\, dz.
\end{equation}
Integrating the above equation from $x_0$ to $a$ and using  the boundary condition $T_a(x_0=a)=0$, we get 
\begin{equation}
T_a(x_0)  =\frac{1}{D} \int_{x_0}^a\, dy\, e^{U(y)/D}\, \int_{-\infty}^y\, e^{-U(z)/D}\, dz.
\label{mean fp}
\end{equation}

For $U(x)=\alpha |x|$, after performing the integrals (for $x_0=0$) we get, 
\begin{equation}
T_a(0)=\frac{2D}{\alpha^2}\left(e^{\alpha a/D}-1\right) -\frac{a}{\alpha}.
\label{Ta-2}
\end{equation}

For the potential given by \eref{potential (i)}, we get
\begin{equation}
T_a(0)= \frac{D}{\alpha^2}\,  \left(e^{\alpha a/D}-1\right) \left[ 2 - e^{-\alpha b/D} +
\sqrt{\frac{\pi}{2}}\, \frac{\alpha}{\sqrt{\mu D}}\, \mathrm{erfc} \left(b \sqrt{\frac{\mu}{2D}}\right)
 \right] -\frac{a}{\alpha}, \quad\text{with}~\mu b= 2\alpha. 
\label{Ta-1}
\end{equation}
On the other hand, 
for the potential given by \eref{potential (iii)}, we get
\begin{equation}
T_a(0)=\frac{D}{\alpha^2} \, \left(e^{\alpha a/D}-1\right) \left[ 2 - e^{-\alpha b/D} + \frac{\alpha b\, e^{-\alpha b/D}}{2 D (\nu-1)}  \right] -\frac{a}{\alpha}.
\label{Ta-3}
\end{equation}
Note that, by taking the limit $b\to \infty$ in both \eref{Ta-1} and \eref{Ta-3}, one gets back \eref{Ta-2}, as required.

\section{Survival probability for a potential given by \eref{linear-quadratic}}

In the main text, we argued that the freezing transition of the decay rate $\theta(a)$ is robust, i.e, it 
occurs for any confining potential $U(x)$ that behaves asymptotically $U(x) \sim |x|$ as $x\to -\infty$. 
The actual value of $a_c$ depends on the details of the potential $U(x)$, but the existence of the 
freezing transition does not depend on the details of $U(x)$ in the bulk as long as $U(x)\sim |x|$ when 
$x\to -\infty$. This argument was based on a general mapping to a quantum problem and using the 
properties of the Schr\"odinger equation in one dimension. We showed that the freezing transition 
coincides with the vanishing of the gap between the ground state (which is a bound state) and the 
continuum of scattering states when $a\to a_c$ from above. We provided an exactly solvable example of 
$U(x)$ in Eqs. (7a) and (7b) of the main text. In this appendix, we provide another example of $U(x)$ 
which differs from the preceding example in the bulk, but still behaves asymptotically $U(x)\sim |x|$ 
when $x\to -\infty$. We show below, analytically and numerically, that the freezing transition again 
occurs, supporting our claim of the robustness of this transition.
We chose the following potential 
\begin{equation}
U(x)=\begin{cases}
\alpha |x| & \text{for}~ x<-b \\[2mm]
\frac{1}{2} \mu  x^2 & \text{for}~ x>-b
\end{cases}
\label{linear-quadratic}
\end{equation}
where $b>0$ and $\mu=2\alpha/b$ for continuity.

As before,  the Laplace transform of the survival probability can be found by solving 
\eref{S-Laplace} piecewise for $x_0 <-b$ and $x_0 >-b$  and then matching the solutions at $x_0=-b$ as well as using the boundary conditions at $x_0=a$ and $x_0\to -\infty$. Since we have already 
solved \eref{S-Laplace} for both linear and quadratic potentials in the examples above, we skip the details here. Moreover, 
for simplicity, we set $\alpha=1$, $D=1$, and $b=1$. We find
\begin{equation}
\tilde{S}_a (s|0) = \frac{1}{s} \bigl[1-\tilde{F}_a(s|0)\bigr]
\end{equation}
where the Laplace transform of the first-passage time  distribution is given by
\begin{equation}
\tilde{F}_a(s|0)=\frac{\sqrt{\pi } e^{-\frac{a^2}{2}} 2^{-\frac{s}{4}}}{\Gamma \left(\frac{s+2}{4}\right)}\, 
\frac{ \chi(s)}{B(s)},
\end{equation}
where
\begin{align}
\chi(s)  &= 2 \sqrt{2} D_{1-\frac{s}{2}}\left(-\sqrt{2}\right)+2 \sqrt{2} D_{1-\frac{s}{2}}\left(\sqrt{2}\right)+\left(\sqrt{4 s+1}+3\right) \left[D_{-\frac{s}{2}}\left(-\sqrt{2}\right)-D_{-\frac{s}{2}}\left(\sqrt{2}\right)\right],\\
B(s) &= \left[2 \sqrt{2} D_{1-\frac{s}{2}}\left(\sqrt{2}\right)-\left(\sqrt{4 s+1}+3\right) D_{-\frac{s}{2}}\left(\sqrt{2}\right)\right] D_{-\frac{s}{2}}\left(\sqrt{2} a\right) \notag\\
&\quad+2 \sqrt{2} D_{1-\frac{s}{2}}\left(-\sqrt{2}\right) D_{-\frac{s}{2}}\left(-\sqrt{2} a\right)+\left(\sqrt{4 s+1}+3\right) D_{-\frac{s}{2}}\left(-\sqrt{2}\right) D_{-\frac{s}{2}}\left(-\sqrt{2} a\right).
\end{align}

By analyzing $\tilde{F}_a(s|0)$ we find that $s=-1/4$ is a branch-point. Moreover, for $a > a_c \approx 
1.06$, $\tilde{F}_a(s|0)$ has a pole at $s^*(a) \in (-1/4,0)$. Therefore, as in \sref{S-I}, here also the 
survival probability behaves as $S_a(t|0) \sim e^{-\theta(a) \, t}$ with $\theta(a)=-s^*(a)$, for $a > 
a_c$ and $S_a(t|0) \sim t^{-3/2}\, e^{-\theta(a)\, t}$ with $\theta(a) = 1/4$, independent of $a$, for $a 
<a_c$.

\begin{figure*}
\includegraphics[width=.45\hsize]{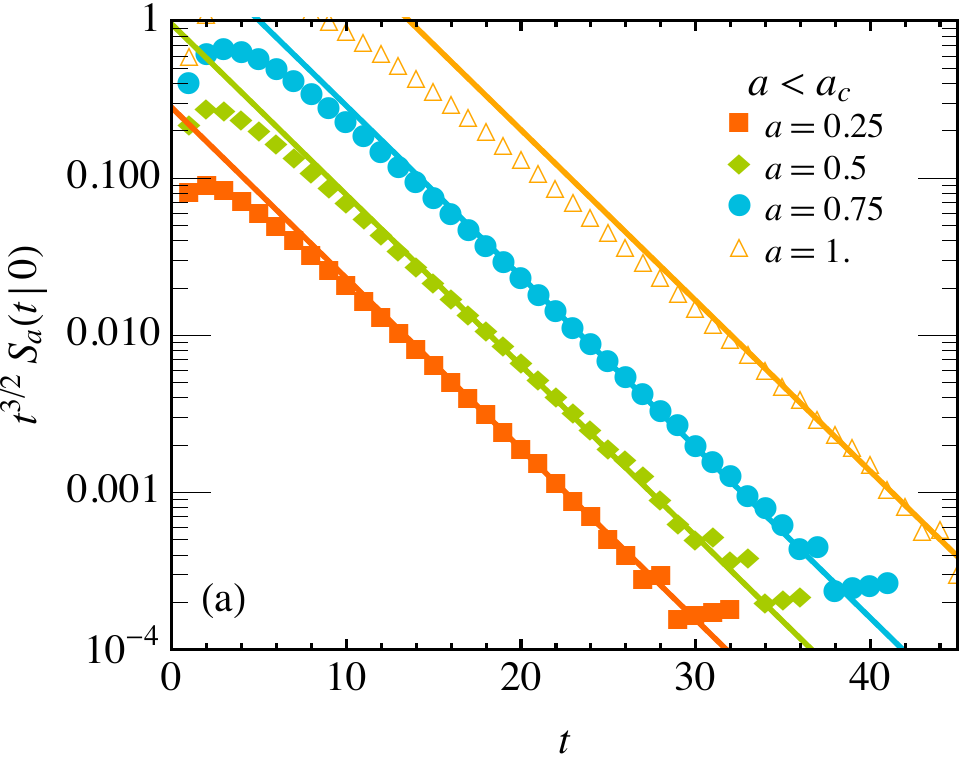}\qquad
\includegraphics[width=.45\hsize]{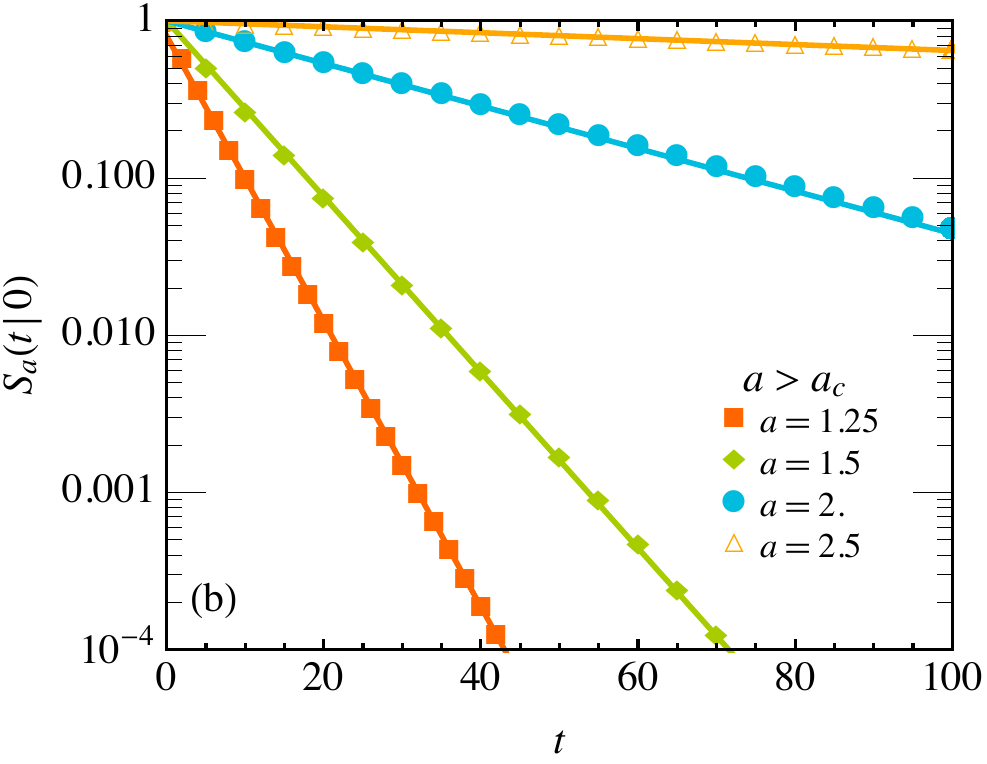}\\[5mm]
\includegraphics[width=.5\hsize]{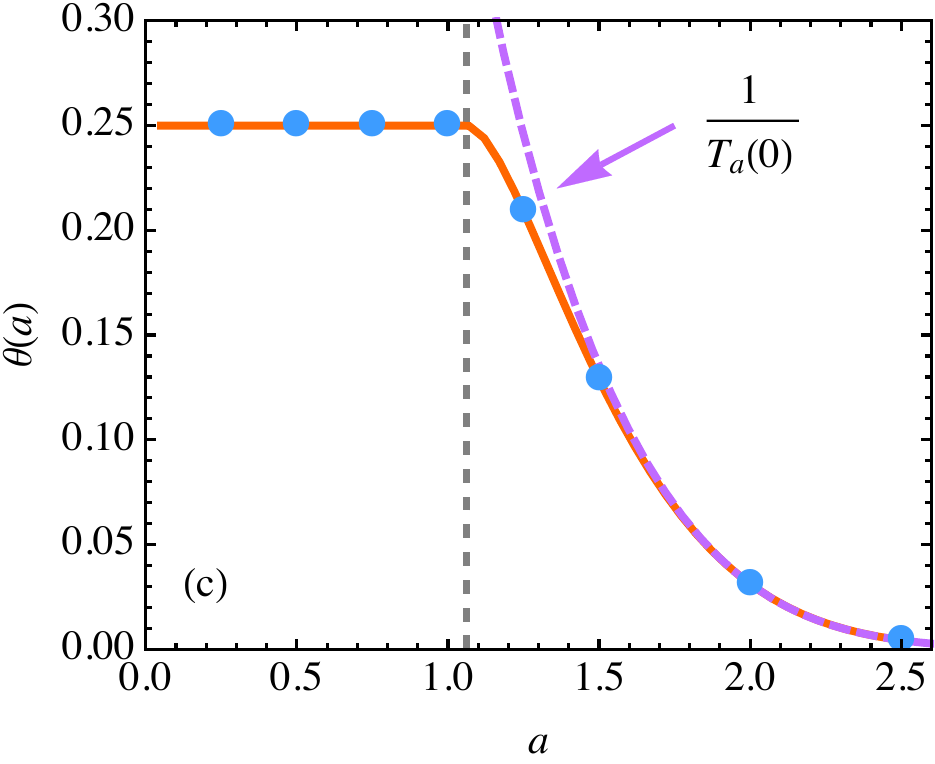}
\caption{\label{surv-fig-lq} (a) and (b): Survival probability of a Brownian particle in a 
potential given by \eref{linear-quadratic}  starting at the position $x_0=0$  
and in the presence of an absorbing barrier at $x=a >0$. We set  $D=1$,  $\alpha=1$, and $b=1$. 
The points are from numerical simulations whereas the solid lines are fit to the exponential 
function $\propto e^{-\theta(a)\, t}$ where the prefactor to the exponential and the 
exponent $\theta(a)$ are chosen to match the simulation points. 
(c) The estimated values of $\theta(a)$ as a function of $a$ are shown by points,  
together with the theoretically computed $\theta(a)$ by the solid (red) line. 
The dashed line shows the inverse of the mean first-passage time calculated exactly 
from \eref{mean fp}. The vertical dashed line marks $a_c\approx 1.06$.}
\end{figure*}

We verify the theoretical predication using numerical simulation. In the presence of an absorbing barrier 
at $x=a$, using the Langevin dynamics, we compute the survival probability of a Brownian particle 
starting from the origin, for various values of $a>0$. The results are shown in \fref{surv-fig-lq}, where 
we find that for larger values of $a$, the survival probability behaves as $S_a(t|0) \sim e^{-\theta(a) 
\, t}$ with a monotonically decreasing $\theta(a)$ as a function of $a$ [see \fref{surv-fig-lq}~(b)]. On 
the other hand, for smaller values of $a$, the survival probability behaves as $S_a(t|0) \sim t^{-3/2}\, 
e^{-\theta(a)\, t}$ with $\theta(a) = 1/4$, independent of $a$ [see \fref{surv-fig-lq}~(a)].  The 
numerically estimated values of $\theta(a)$ together with the theoretically calculated values are shown 
in \fref{surv-fig-lq}~(c), which again shows the freezing transition similar to the one reported in the 
main text of the Letter.

\section{Survival probability for a potential given by \eref{sqrt-linear}}

\begin{figure*}
\includegraphics[width=.45\hsize]{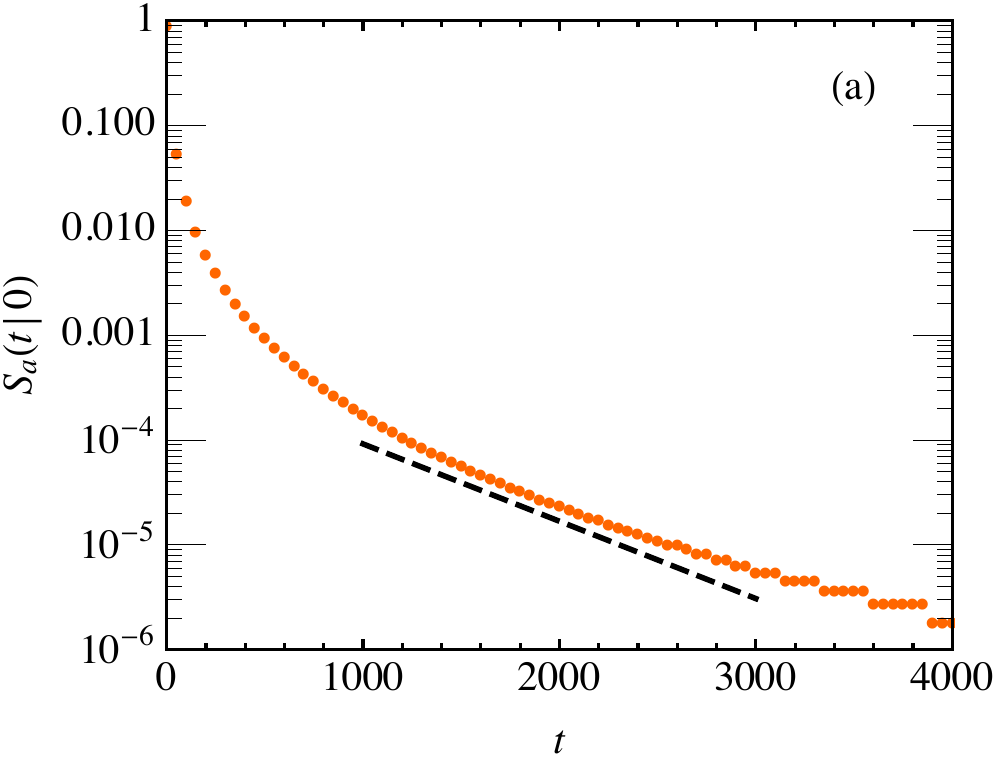}\qquad
\includegraphics[width=.45\hsize]{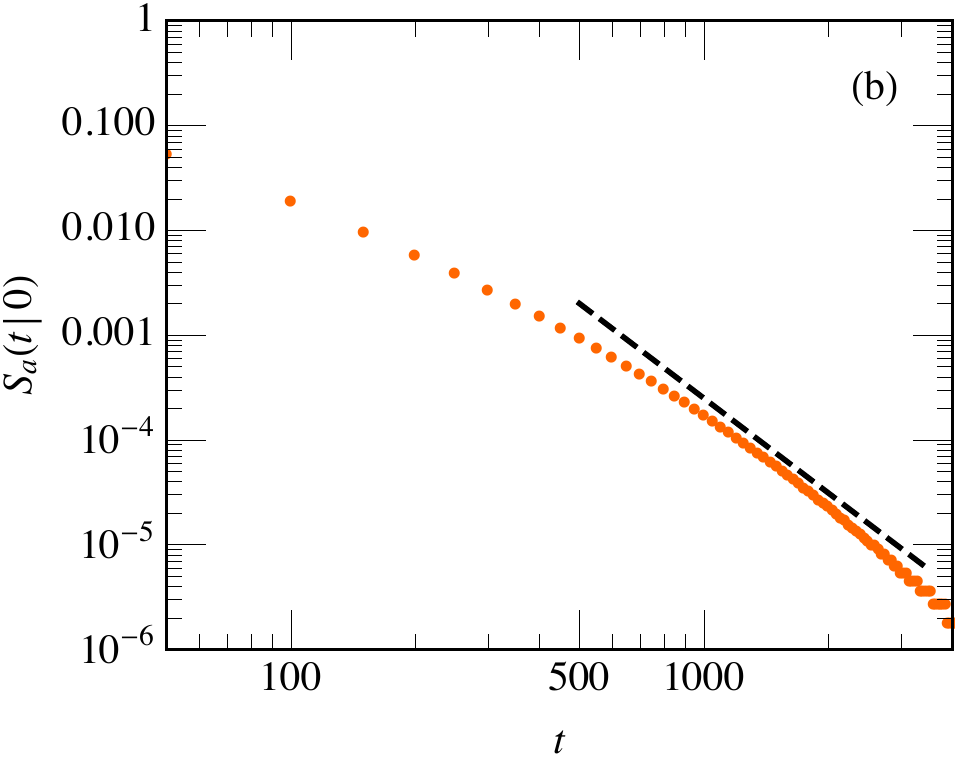}\\[5mm]
\includegraphics[width=.5\hsize]{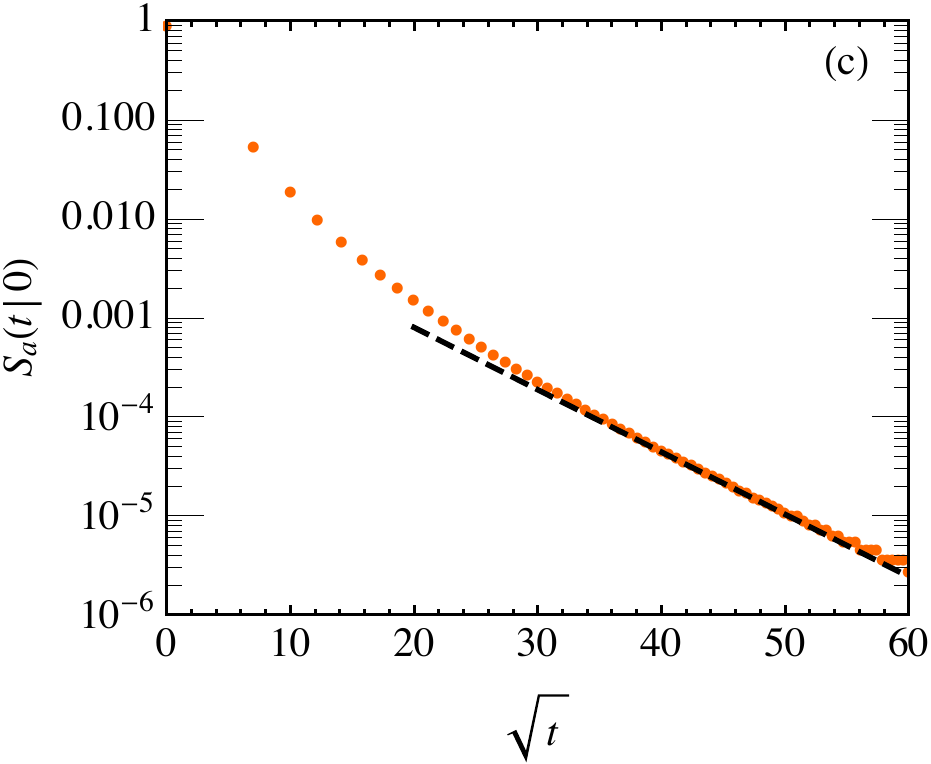}
\caption{\label{surv-fig-sqrt} Survival probability of a Brownian particle in a 
potential given by \eref{sqrt-linear}  starting at the position $x_0=0$  
and in the presence of an absorbing barrier at $x=a >0$. We set  $D=1$,  $\alpha=1$,  $b=1$, and $c=1/2$.  The points in (a)-(c) are the same numerical simulation data for the survival probability $S_a(t|0)$ presented in different scales whereas the dashed lines represent different functions.  (a)  The dashed line  indicates that $S_a(t|0)$ has a slower than an exponential $\propto e^{-\theta t}$ decay, where $\theta$ and the proportionality constants are fitting parameters. (b) The dashed line indicates that $S_a(t|0)$ has a faster than a power-law $\propto t^{-3}$ decay. (c) Semi-log plot $S_a(t|0)$ as a function of $\sqrt{t}$. The dashed line demonstrates a stretched exponential decay $S_a(t|0) \sim e^{-r \sqrt{t}}$, where $r$ is a fitting parameter. 
 }
\end{figure*}

In the main text, using a mapping to a quantum problem, we argued that if the confining potential $U(x)$ increases slower than $|x|$ as 
$x\to -\infty$, then $\theta(a)=0$ for all $a$, indicating a slower than exponential decay with time, of the 
first-passage/survival  probability. We have analytically  demonstrated that for $U(x) \sim \ln (-x)$ as $x\to -\infty$, both the first-passage and the survival probability exhibit power law decay with time  [see \sref{logx-potential-left}]. Here we consider another case
\begin{equation}
U(x)=\begin{cases}
2 c \sqrt{-x} & \text{for}~ x<-b \\[2mm]
\alpha |x| & \text{for}~ x>-b
\end{cases}
\label{sqrt-linear}
\end{equation}
 where $b >0$,  and for the sake of continuity of the potential at $x=-b$, we set $c=\alpha \sqrt{b}/2$. Using numerical simulation we show [see \fref{surv-fig-sqrt}] that the survival probability again has a slower than an exponential decay, namely, $S_a(t|0) \sim e^{- r \sqrt{t}}$.